\documentclass[preprint,5p,twocolumn,number]{elsarticle}

\usepackage{lineno,hyperref}
\modulolinenumbers[5]

\journal{The Journal of Systems and Software}

\usepackage{amsmath,amssymb,amsfonts}
\usepackage{algorithmic}
\usepackage{graphicx}
\usepackage{textcomp}

\newtheorem{definition}{Definition}[section]

\usepackage{float}
\usepackage[table,xcdraw,dvipsnames]{xcolor}
\usepackage{subcaption}
\usepackage{capt-of}
\usepackage[hang]{footmisc}
\usepackage{listings}
\usepackage{color}

\makeatletter
\def\lst@makecaption{%
    \def\@captype{table}%
    \@makecaption
}
\makeatother

\usepackage{paralist}
\usepackage{csquotes}

\usepackage{pifont}

\usepackage{multicol}
\usepackage{multirow}
\usepackage[flushleft]{threeparttable}

\usepackage{tabularx,booktabs}
\newcolumntype{b}{X}
\newcolumntype{s}{>{\hsize=.4\hsize}X}
\newcolumntype{C}{>{\centering\arraybackslash}X}

\usepackage{stfloats}

\PassOptionsToPackage{hyphens,spaces,obeyspaces}{url}
\usepackage{hyperref}

\usepackage[capitalise,nameinlink,noabbrev]{cleveref}

\newcommand\req[1]{\mbox{Req.~#1}}

\begin{document}

\begin{frontmatter}

\title{On the Serverless Nature of Blockchains and Smart Contracts}

\author{Vladimir Yussupov}
\ead{yussupov@iaas.uni-stuttgart.de}

\author{Ghareeb Falazi}
\ead{falazi@iaas.uni-stuttgart.de}

\author{Uwe Breitenb{\"u}cher}
\ead{breitenbuecher@iaas.uni-stuttgart.de}

\author{Frank Leymann}
\ead{leymann@iaas.uni-stuttgart.de}

\address{Institute of Architecture of Application Systems, University of Stuttgart, Germany}


\begin{abstract}
Although historically the term serverless was also used in the context of peer-to-peer systems, it is more frequently associated with the architectural style for developing cloud-native applications.
From the developer's perspective, serverless architectures allow reducing management efforts since applications are composed using provider-managed components, e.g., Database-as-a-Service~(DBaaS) and Function-as-a-Service~(FaaS) offerings.
Blockchains are distributed systems designed to enable collaborative scenarios involving multiple untrusted parties.
It seems that the decentralized peer-to-peer nature of blockchains makes it interesting to consider them in serverless architectures, since resource allocation and management tasks are not required to be performed by users.
Moreover, considering their useful properties of ensuring transaction's immutability and facilitating accountable interactions, blockchains might enhance the overall guarantees and capabilities of serverless architectures.
Therefore, in this work, we analyze how the blockchain technology and smart contracts fit into the serverless picture and derive a set of scenarios in which they act as different component types in serverless architectures.
Furthermore, we formulate the implementation requirements that have to be fulfilled to successfully use blockchains and smart contracts in these scenarios.
Finally, we investigate which existing technologies enable these scenarios, and analyze their readiness and suitability to fulfill the formulated requirements.
\end{abstract}


\end{frontmatter}


\section{Introduction}
\label{sec:intro}

Modern cloud-native application development relies on a plethora of available cloud service models that allow flexibly making the trade-off between out-of-the-box integration and control over the infrastructure.
For example, developers can decide to have more control over the infrastructure using Infrastructure-as-a-Service~(IaaS) offerings, or instead they can delegate some of their management responsibilities using more provider-managed cloud service models such as Platform-as-a-Service~(PaaS).
While being also used in contexts such as peer-to-peer (P2P) networks~\cite{serverless:fox2017status}, the notion of \textit{serverless computing} is more often associated with cloud computing and the Function-as-a-Service~(FaaS) cloud service model in particular~\cite{serverless:jonas2019cloud}.
Essentially, FaaS serves as one of the main enablers for hosting business logic in serverless architectures: developers are able to host arbitrary, event-driven code snippets that are managed by cloud providers and charged on a per-invocation basis~\cite{serverless:baldini2017serverless}.
Additionally, multiple providers offer so-called \textit{function orchestrators} that allow specifying and executing function workflows that orchestrate several individual functions, e.g., AWS Step Functions~\cite{aws:step-functions}.
Interestingly, these function orchestrators are used in a serverless manner too, i.e. the function implementations themselves as well as their orchestrations can be developed \textit{serverlessly}.
Hence, developers can combine event-driven functions, function orchestrators, and different serverless event emitters, e.g., Database-as-a-Service (DBaaS) offerings or messaging services, to implement serverless architectures with a stronger focus on business logic and component integration instead of infrastructure management.

Although the term \enquote{serverless} is often defined ambiguously, e.g., linked only with FaaS, its main idea revolves around several characteristics, namely (i) the decoupling of computation and storage, (ii) resource allocation being managed by a third party, and (iii) per-usage pricing models~\cite{serverless:jonas2019cloud}.
Apart from FaaS and function orchestrators being prominent examples of \textit{serverless component types}, there are many other provider-managed component types that follow these serverless characteristics, e.g., object storages, NoSQL databases, or messaging and streaming platforms when offered \enquote{as a service}.
Hence, this variety of component types enables implementing complex \textit{serverless architectures} comprising only serverless components, e.g., functions hosted on a FaaS platform interact with tables in a hosted NoSQL database, etc.
Moreover, multiple serverless components can play the role of \textit{event emitters} that are able to trigger FaaS functions or function workflows, e.g., a database insert event triggers a single function, or an entire function workflow that analyzes the inserted data.
Despite the considerable amount of available serverless component types, the boundaries of \emph{\enquote{serverless-ness}} are not fixed and can be extended by other component types fitting into the serverless paradigm~\cite{serverless:baldini2017serverless, serverless:jonas2019cloud}.

On the other hand, \textit{blockchains} are distributed systems designed to facilitate the collaboration of untrusting participants without depending on third-parties.
This is achieved by innovatively solving the problem of consensus in the presence of Byzantine adversaries~\cite{Barborak1993} in a P2P setup lacking any centralized elements, and ensuring the practical immutability of data stored in these systems.
Early blockchains, such as Bitcoin~\cite{Nakamoto2008}, only supported the atomic transfer of tokens, known as \textit{cryptocurrencies}, from one account to another.
However, since the introduction of \textit{smart contracts} by Ethereum~\cite{wood2018ethereum}, blockchains have expanded their applicability beyond cryptocurrencies to domains such as supply chain management, healthcare, IoT, or data management~\cite{Casino2019}.
Smart contracts are applications that run and store their state on the blockchain, thus equipping external applications integrating with them with trustworthy, deterministic on-chain functionalities, such as payment, logging, and state management.
A \textit{smart contract} is invoked by sending a cryptographically signed request, i.e., a \textit{blockchain transaction}, holding the required invocation arguments to any node of the P2P system~\cite{wood2018ethereum}.

Surprisingly, the core characteristics of serverless application components, i.e., decoupling of computation and storage, third-party-managed resource allocation and per-usage pricing models, to a certain degree seem to be also related to blockchains and smart contracts due to their decentralized P2P nature.
Furthermore, serverless applications comprising various provider-managed components might suffer from trust issues in certain cases.
For example, developers might require guarantees that functions hosted on FaaS platforms are executed as intended, e.g., triggered only once by the exact designated event, or that the input parameter values sent to it are not manipulated.
Moreover, the integration of multi-application or multi-provider serverless architectures also entails trust issues similar to the ones commonly solved by blockchains.
As a result, it seems very appealing \textit{to employ blockchains as components in serverless architectures to enhance the trust guarantees and integration capabilities}.

In this work, we investigate this hypothesis by analyzing how blockchains and smart contracts can be employed as components in serverless architectures.
First, we present a detailed analysis of scenarios in which blockchains play different roles in serverless architectures, e.g., blockchains as event emitters, smart contracts as FaaS functions, and blockchains for function orchestration.
Secondly, we analyze how blockchains can help in serverless cross-application or cross-platform integration scenarios, i.e., scenarios in which multiple serverless components hosted in the same or different cloud(s) need to interact with each other in an accountable way.
As a result of these investigations, we identify the challenges and formulate the requirements that have to be fulfilled by the employed technology stack to enable using blockchains in serverless architectures.
Finally, we analyze which existing technologies can serve as enablers for the identified scenarios and determine their ability to fulfill the formulated implementation requirements.
To summarize, the main contributions of this work are threefold.
\begin{enumerate}[(i)]
	\item We analyze how blockchains and smart contracts can be used as components in serverless architectures.
	\item We formulate the implementation requirements supporting usage of blockchains in these scenarios.
    \item We identify the existing technologies enabling these scenarios and analyze how capable are they to fulfill the detected implementation requirements.
\end{enumerate}

The remainder of the paper is structured as follows.
\Cref{sec:bg} presents the necessary background and motivation.
\Cref{sec:scenarios} discusses how blockchains can be used in serverless architectures.
\Cref{sec:requirements} formulates the requirements for implementing the mentioned scenarios.
\Cref{sec:system-support} presents an analysis of existing technologies with respect to their readiness to fulfill the identified requirements and discusses the key highlights.
\Cref{sec:blockchain-awareness} elaborates on the core findings of this work motivating the future research directions.
\Cref{sec:related-work,sec:conclusion} discuss the related publications and conclude this work.


\section{Background and Motivation}
\label{sec:bg}

This section provides the necessary details on serverless architectures, blockchains, and smart contracts to facilitate the understanding of the remainder of this work.
Moreover, we present the motivation and problem statement that are concluded by our main research questions.

\subsection{Serverless Architectures}
In the cloud context, the term \textit{serverless} is often associated with FaaS, a particular \textit{serverless component type} that allows hosting business logic in serverless architectures~\cite{serverless:jonas2019cloud,fowler:serverless}.
With FaaS, developers are able to deploy custom code that can be triggered by events originating from multiple provider-managed cloud service offerings.
The underlying FaaS platforms, such as Azure Functions~\cite{ms:azure-functions} or AWS Lambda~\cite{aws:lambda}, are managed by cloud providers. 
More specifically, these platforms are responsible for automatic resources allocation and scaling idle instances to zero, which eliminates the need to pay for unused resources.
The idea of using FaaS functions found its applicability in multiple real-world scenarios, e.g., serverless APIs or event-driven data pipelines~\cite{serverless:baldini2017serverless,serverless:jonas2019cloud}.
Moreover, FaaS functions can be orchestrated using another type of serverless components, namely \textit{function orchestrators} such as AWS Step Functions~\cite{aws:step-functions}.

However, it is important to point out here that, in the majority of cases, designing a serverless architecture using only FaaS functions and function orchestrators is not sufficient.
In fact, one of the main advantages of FaaS is the out-of-the-box integration with a plethora of provider-managed services such as databases, messaging, logging, and monitoring services that can serve as \textit{event emitters} triggering FaaS functions~\cite{related:motivation:baldini2017serverless}.
Moreover, the provider might also offer dedicated event hub services that are responsible for aggregating and distributing various event types also originating from other cloud infrastructures and SaaS offerings.
For example, FaaS-based serverless architectures are an appealing option for implementing Extract-Transform-Load pipelines, with different independent functions~(or even function workflows) being responsible for various sub-tasks in each phase, e.g., implementing different transformation functions triggered by events originating from heterogeneous data sources~\cite{ms:serverless-examples}.
Therefore, we call such composition of multiple distinct serverless components a \textit{cloud-native serverless architecture}, which allows developers to focus on business logic composition instead of infrastructure management tasks.
Thus, serverless component types and the ways they can be composed are already quite heterogeneous: various storage and messaging solutions, event-driven flows and function orchestration can be combined with special-purpose services, such as AWS Alexa~\cite{aws:alexa} or IBM Watson~\cite{ibm:watson}.
Nevertheless, the boundaries of \enquote{\emph{serverless-ness}} are not fixed, and emerging as well as existing concepts can be used as a basis for new serverless component types in the future~\cite{related:motivation:baldini2017serverless,serverless:jonas2019cloud}.

Based on the idea of severless characteristics by Jonas et al.~\cite{serverless:jonas2019cloud}, we define the terms "Serverless Component" and "Serverless Architecture" for this article as follows. 
\begin{definition}[Serverless Component]
A serverless component is a (cloud) application's building block which possesses the following characteristics:
\begin{compactenum}[(i)]
    \item Storage and computation can be decoupled, i.e., scaling, provisioning, and pricing happen separately.
    \item Resource allocation is managed by a cloud provider.
    \item Charging is associated with the active usage of component, i.e., costs do not involve idle time charges.
\end{compactenum}
\end{definition}

\begin{definition}[Serverless Architecture]
A serverless architecture is a cloud-native architecture that comprises only serverless components.
\end{definition}

\subsection{Blockchains and Smart Contracts}
\textit{Blockchains} can be abstractly described as distributed systems that allow independent parties to conduct collaborative processes even while having limited mutual trust.
Early blockchain systems, such as Bitcoin~\cite{Nakamoto2008}, supported simple transactional applications that allow users to exchange \textit{cryptocurrencies}, which are virtual tokens of value managed in a decentralized manner by the collective blockchain system in a way that prevents the problem of double spending~\cite{Chohan2017DoubleSpending}.
Later, more sophisticated applications were supported via the introduction of smart contracts~\cite{wood2018ethereum,Szabo1996SmartContracts}.
\textit{Smart contracts} are user-defined transactional applications stored on the blockchain and guaranteed to be executed the way they are programmed.
Therefore, they are suitable to implement sophisticated business use-cases, which involve untrusting partners.
To invoke a smart contract, a client application formulates a cryptographically signed request message, known as a \textit{blockchain transaction}, that contains entries like the unique address of the smart contract, the signature of the specific function to be invoked, as well as the arguments passed to it, and sends it to one of the network peers, known as \textit{blockchain nodes}.
Then the node validates the transaction, and includes it in a distributed, blockchain-specific \textit{consensus mechanism} that ensures all honest peers agree on the contents of the transaction and its global order among other transactions.
The result of this mechanism is a \textit{block} of transactions that is appended at the end of a fully replicated list of blocks, which are chained by their hashes.
This list is known as the \textit{blockchain data structure}, and the way it is organized, in addition to the consensus mechanism, both ensure that the stored transactions are practically immutable~\cite{Nakamoto2008}.
Before the next block is added to the blockchain, all nodes execute the transactions of the latest block sequentially, which involves invoking the corresponding smart contract functions, and deterministically updating the state managed by them~\cite{wood2018ethereum}.
Since blockchain transactions are cryptographically signed by their initiators, and immutably stored in the replicated data structure, the \textit{non-repudiation} of actions performed by blockchain clients is ensured~\cite{cachin2017consensus}.
Applications whose business logic is managed by one or more smart contracts are known as \textit{decentralized applications}, or DApps~\cite{Buterin2014EthereumWhitePaper}.

Basically, there exist two major categories of blockchains.
Blockchains like Bitcoin and Ethereum~\cite{wood2018ethereum} are \textit{permissionless (or public)} in the sense that accessing the system in any role is open for anyone.
These systems favor absolute decentralization at the cost of having relatively weak privacy and performance capabilities.
\textit{Permissioned (or private) blockchains} were introduced as an alternative that guarantees data confidentiality, and generally ensures better performance~\cite{cachin2017consensus,Yaga2018BCTechReview}.
However, these desirable properties influence the degree of decentralization, since joining the system becomes under the control of a single entity.
Therefore, we see that \textit{there is no single blockchain technology that is capable of satisfying all relevant use-cases}, which means that existing and new variations of blockchains would continue to co-exist~\cite{falazi2019permissioned}.

\subsection{Motivation and Problem Statement}
\label{sec:motivation}
As mentioned previously, the early idea of a serverless computation was connected with P2P systems in general, with the main emphasis on the reduced management efforts and the weaker role of traditional hosting components referred to as servers.
Interestingly, blockchains, being distributed and decentralized systems~(in the majority of cases), can also be seen as possible serverless component types as they are not managed in a traditional sense, and often allow executing custom code in the form of smart contract functions.
More specifically, to execute a smart contract function, a client application only needs to submit a request message, i.e., a blockchain transaction, to a blockchain node.
The node itself does not need to be a part of the client application, hence, its management is transparent to application developers.
The blockchain protocol is responsible for scheduling and executing designated functions on behalf of client applications, which resembles the FaaS platform's behavior. 
Hence, it seems that there exist several possibilities for employing blockchains~(and smart contracts) as components in serverless architectures.
Considering the guarantees brought by blockchains, a possibility to use them as serverless components can also help extending the applicability and trust guarantees of serverless architectures in general.
However, this requires understanding which roles blockchains and smart contracts can play in serverless architectures.
For instance, serverless architectures are often reactive since they rely on event-driven FaaS functions.
In theory, events triggering FaaS functions can also originate from blockchains.
Moreover, smart contract functions and blockchains as a whole can be considered in scenarios where they are used as substitutes for FaaS functions and function orchestrators.
Thus, the aforementioned examples already demonstrate several distinct roles blockchains and smart contracts could play in serverless architectures.
Therefore, the research questions of this work can be formulated as follows: 
\begin{center}
    \noindent
    \setlength{\fboxsep}{2mm}
    \fbox{
        \centering
        \begin{minipage}{.9\linewidth}
            \begin{compactenum}
                \item[\textit{\textbf{RQ1:}}] \textit{\enquote{In which scenarios can blockchains and smart contracts be used in serverless architectures and which components can they replace?}}
                \item[\textit{\textbf{RQ2:}}] \textit{\enquote{What are the technical requirements for implementing such scenarios?}}
                \item[\textit{\textbf{RQ3:}}] \textit{\enquote{Which existing technologies can facilitate fulfillment of these requirements?}}
            \end{compactenum}
    \end{minipage}}
    \setlength{\fboxsep}{2mm}
   \vspace{.2\baselineskip}
\end{center}


\section{Blockchains in Serverless Architectures}
\label{sec:scenarios}

As discussed in~\Cref{sec:bg}, blockchains and smart contracts share similarities with the concept of a serverless component, which opens up different opportunities for using them in serverless architectures.
Firstly, since events that trigger FaaS functions and function workflows are mainly emitted by other third-party services supported by the underlying FaaS platform, one possible option that can be investigated is how to use \textit{blockchains as event emitters} triggering parts of a given serverless architecture.
Next, when compared to FaaS functions, smart contracts have several interesting property overlaps, which unfolds other scenarios, e.g., using \textit{smart contracts as FaaS functions}, or \textit{blockchains and smart contracts as actors in serverless workflows}.
Moreover, all these examples can also be considered not in the scope of a single application, but rather as a way to \textit{enable the integration of multiple serverless applications}, possibly hosted on different clouds.

In this section, we present our first contribution in which we analyze the scenarios of using blockchains and smart contracts are used as different serverless component types, and discuss which technical implications have to be considered.
In particular, we discuss the following four scenarios:
\begin{compactenum}
    \item [\textit{\textbf{Scenario-1:}}] blockchains are used as serverless event emitters~(discussed in~\cref{sec:blockchains-as-event-emitters}),
    \item [\textit{\textbf{Scenario-2:}}] smart contracts are used as serverless functions~(discussed in~\cref{subsec:scenario-3}),
    \item [\textit{\textbf{Scenario-3:}}] blockchains are used for function orchestration~(discussed in~\cref{sec:scenario-function-orchestration}), and 
    \item [\textit{\textbf{Scenario-4:}}] blockchains are used as facilitators for the integration of serverless applications~(\cref{sec:integration}).
\end{compactenum}

\noindent
Each scenario is discussed following the standardized structure, which comprises four parts:
\begin{inparaenum}[1)]
    \item \textit{Motivation and Analysis of Serverless Properties},
    \item \textit{Goal},
    \item \textit{Realization}, and
    \item \textit{Pitfalls to Consider}.
\end{inparaenum}
In the first part, the context in which the scenario is set and the major motivation behind it are discussed together with an analysis of relevant serverless properties of blockchains, which make them interesting for this particular case.
For example, the ability to emit events makes blockchains interesting for usage as serverless event emitters.
Moreover, this part discusses how the usage of blockchains enhances the guarantees provided by serverless architectures.
In the second part, we formulate the main goal of the scenario.
The last two parts provide a technical discussion about realization strategies and pitfalls that need to be considered.

\subsection{Scenario 1: Blockchains as Serverless Event Emitters}
\label{sec:blockchains-as-event-emitters}

In the first scenario, we focus on the possibility to use blockchain events in serverless architectures.

\subsubsection{Motivation and Analysis of Serverless Properties}
Serverless architectures often employ the event-driven computing paradigm, allowing triggering FaaS functions using events originating from multiple provider-managed services.
Smart contracts can be programmed to emit events when certain conditions are met while executing their code.
For example, a smart contract managing the provenance of a seafood product may be designed to emit an event when the product is moved from one handler to another, e.g., from the food processor to the end-distributor~\cite{Falazi2020unified}.
In addition, apart from application-specific events, blockchains can also emit other kinds of events interesting to external applications, e.g., events signifying (i) a successful processing and inclusion of a blockchain transaction, or (ii) the mining of a new block and appending it to the blockchain data structure.

\subsubsection{Goal}
The ability to emit different kinds of events opens many opportunities for employing blockchains as serverless event emitters.
This scenario focuses on \textit{employing blockchains in serverless architectures as serverless event emitters that are able to trigger functions hosted on FaaS platforms}.

\subsubsection{Realization}
\begin{figure}[t!]
	\centering
	\includegraphics[width=.95\columnwidth]{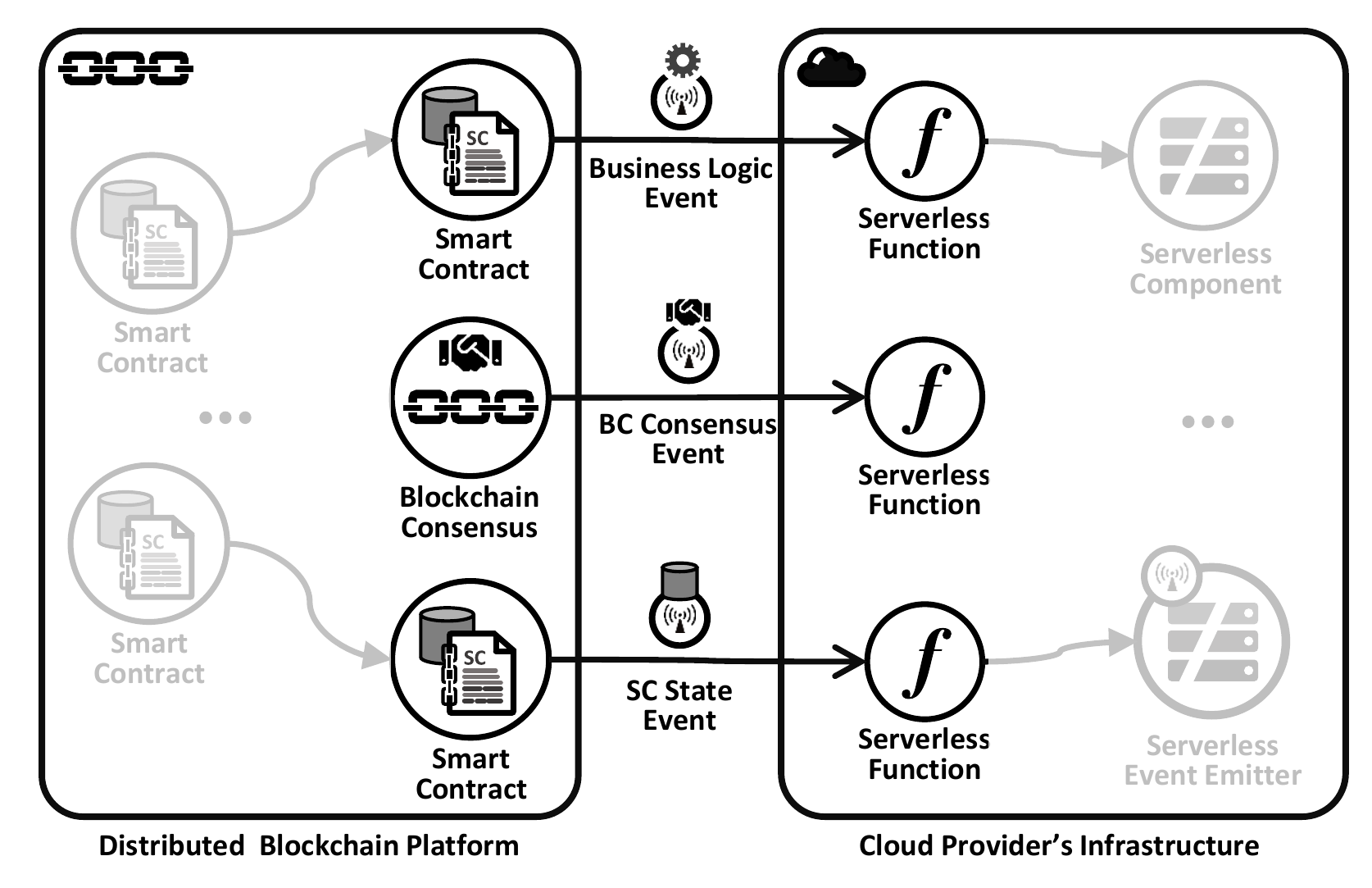}
	\caption{Blockchain as a serverless event emitter.}
	\label{fig:blockchain-event-emitter}
\end{figure}

FaaS functions are either invoked synchronously in direct calls issued, for example, by other functions, or asynchronously via events that the functions are pre-configured to consume.
Specific platform-managed services, such as message queues, event streaming platforms, or database management systems, are typical sources of such events.
Thus, this scenario describes how blockchain events originating from application-specific components, e.g., smart contracts, or system-wide services, e.g., the consensus mechanism, can be used  to asynchronously trigger FaaS functions as shown in \cref{fig:blockchain-event-emitter}.
Specifically, smart contracts can be utilized to embed the business logic of certain use-cases.
In such situations, they can emit events that signal reaching specific business conditions in their logic, i.e., \textit{business logic events}.
On the other hand, smart contracts are often used to store and manage data with a predefined structure.
This is, for example, the case with ERC-20 smart contracts that manage most of Ethereum user-defined tokens~\cite{Fenu2018ICO}.
Here, the data itself is part of the permanent state of the smart contract, whereas the functions of the smart contract handle the data management, and they may be programmed to emit events that indicate to external applications that the corresponding functions were invoked causing the state to be changed.
We will refer to these events as \textit{smart contract state events}.
Finally, as we mentioned earlier, blockchains emit system-wide events that are mainly meant to inform external applications about conditions related to the underlying consensus mechanism, such as the creation of a new block, or the change of the set of validators (in the case of permissioned blockchains), etc.
We will refer to these events as \textit{blockchain consensus events}.
As shown in \cref{fig:blockchain-event-emitter}, these dedicated event types can be used to asynchronously trigger the execution of FaaS functions.
Depending on the chosen infrastructure's capabilities, such triggering might even be supported natively, e.g., blockchain events are captured by a monitoring component and passed to the FaaS platform, e.g., Microsoft Azure supports such integration as will be discussed in-detail in~\cref{sec:system-support}.

\subsubsection{Pitfalls to Consider}
The nature of blockchain systems requires the execution of  smart contracts to be deterministic, i.e., executing the same smart contract function using the same request parameters with the same blockchain global state should always yield the same results, even years in the future.
This means that the code of the smart contract cannot directly involve any sort of randomness, even indirect randomness resulting from communicating with external systems, since they might fail or might return different results to the smart contract in different executions.
That is why blockchains cannot actively access the external world~\cite{Al-Breiki2020}.
In our scenario, this imposes strict requirements on implementing the interaction from the blockchain to a serverless function (shown in \cref{fig:blockchain-event-emitter}), i.e., the blockchain is not able to actively deliver~(push) the custom events to FaaS functions.
In general, blockchain events resulting from executing certain transactions are stored in a dedicated section within the block that contains these transactions, and since the blocks are publicly accessible, external applications periodically read these logs~(pull) searching for events relevant for them.
This means that a similar mechanism needs to be used for delivering such events to FaaS functions.
This can be done \enquote{natively} at the provider level by having a dedicated component that polls the logs and informs interested FaaS functions when relevant events are detected, or it can be done \enquote{manually} at the application level, for example, by leveraging a scheduled function trigger, such as the Azure Timer trigger~\cite{ms:azure-timerTrigger}, that periodically fires off the execution of a special serverless function that only has the purpose of querying the blockchain log for relevant events and triggering a second function that holds the business logic if such events are found.

\subsection{Scenario 2: Smart Contracts as Serverless Functions}
\label{subsec:scenario-3}
This scenario analyzes how smart contracts can be used as an alternative to functions hosted on FaaS platforms.

\subsubsection{Motivation and Analysis of Serverless Properties}
From a high-level view, smart contracts are similar to FaaS functions as both share the following characteristics:
\begin{compactitem}
    \item \textit{Per-invocation pricing:}
    from the client application's perspective, the only cost of executing a smart contract function is the fees potentially embedded within request messages.
    No additional costs are incurred, e.g., due to managing servers or other resources.
    In this sense, smart contract functions are priced per-invocation, similar to FaaS functions.
    \item \textit{Scaling to zero:}
    blockchain smart contracts do not need to be running all the time.
    In fact, between invocations they always scale down to zero since the execution of smart contract functions is (i) inherently short-lived because of its high cost and (ii) stateless in the sense that local variables are not shared among the invocations.
    Of course, smart contracts can access a state stored permanently in the blockchain data structure itself, but this is comparable to storing data in a database, and does not mean that they are stateful. Thus, smart contracts are similar to FaaS functions.
    \item \textit{Hosting custom code:}
    most blockchain smart contracts support general-purpose Turing-complete programming languages~\cite{bartoletti2017_SmartContractAnalysis} including Solidity, Go, Python, and JavaScript.
    From a functional viewpoint this means that arbitrary logic can be encoded within smart contracts, much like FaaS functions.
\end{compactitem}
Considering these similarities, it seems feasible to use smart contracts as an alternative for FaaS functions.
Moreover, despite the fact that smart contract function executions are generally short, expensive, and can only manipulate small amounts of data~\cite{Kim2018}, they provide certain guarantees that lack from their FaaS counterparts.
For example, smart contract functions only process cryptographically signed request messages that are immutable and permanently stored.
This helps solving many trust-related issues on multiple levels.
Furthermore, smart contracts run on the decentralized infrastructure of the blockchain, and, as we saw earlier, allow emitting verifiable events when certain situations defined in their code occur.
These events cannot be faked and depend solely on their inputs and the logic encoded into the smart contract code.

\subsubsection{Goal}
Overall, since security issues are among highly-researched questions related to FaaS platforms~\cite{Yussupov2019_SystematicMappingStudy} and taking blockchain properties into account, in certain cases, smart contracts can be considered as an alternative to FaaS functions with blockchains serving as \textit{trust-enhanced execution environments}.
Therefore, this scenario focuses on \textit{how to employ smart contracts as an alternative to FaaS functions in serverless architectures when certain security aspects play an important role.}

\begin{figure}[bp!]
    \centering
    \includegraphics[width=.95\columnwidth]{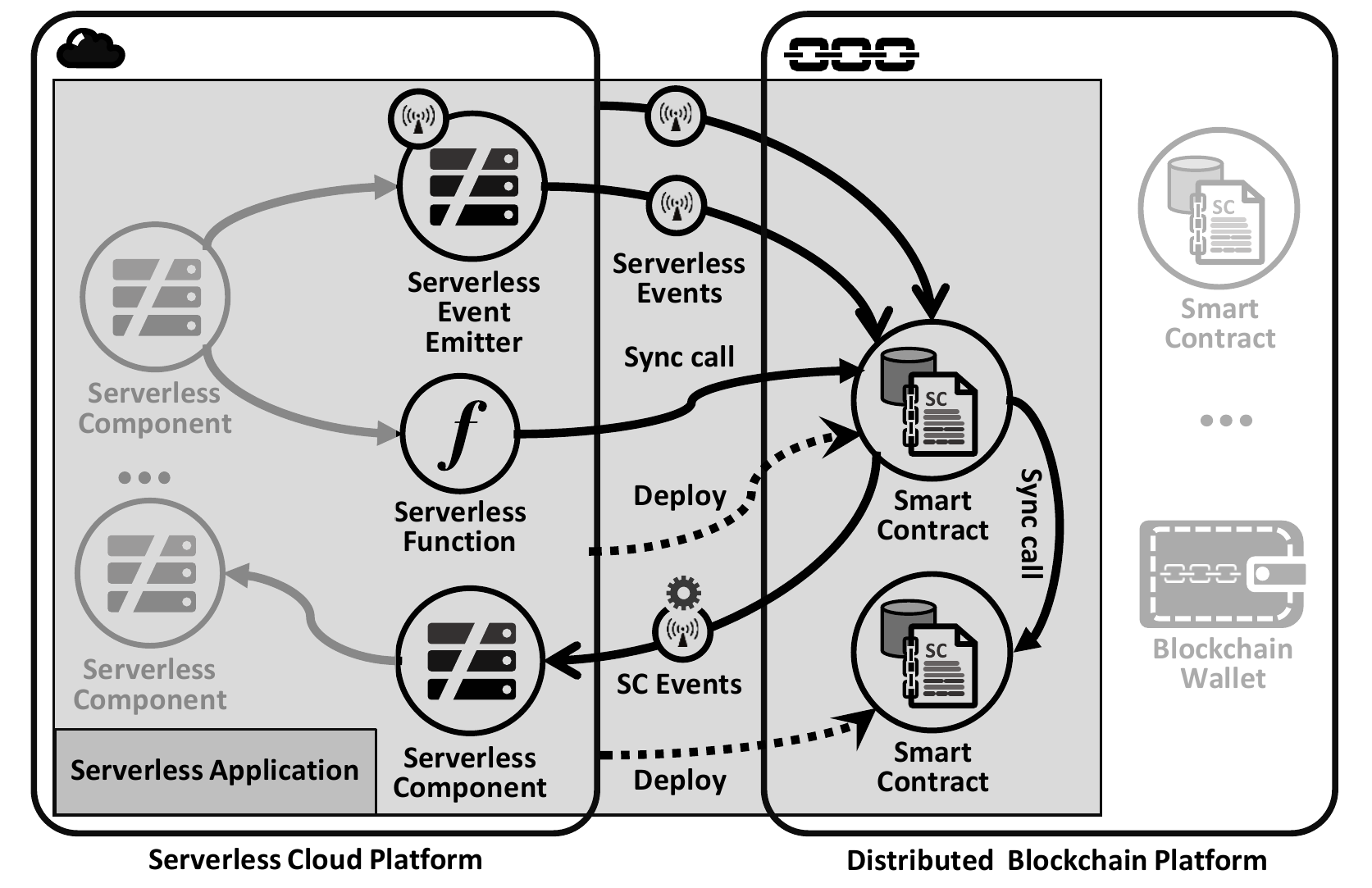}
    \caption{Smart contracts as serverless functions triggered by synchronous/asynchronous calls, and triggering other components.}
    \label{fig:smart-contracts-as-functions}
\end{figure}

\subsubsection{Realization}
As depicted in \cref{fig:smart-contracts-as-functions}, smart contract-enabled blockchains can be seen as trust-enhanced function execution environments in the following ways:
\begin{compactenum}[(i)]
    \item A smart contract can be used either explicitly by serverless applications or transparently by serverless platforms to immutably and permanently store application logs.
    This ensures that program events can be verifiable and attributed to the point in time in which they occurred.
    This is especially helpful to log certain user interactions with the application so that accountability~\cite{Kuesters2010_Accountability} is guaranteed.
    \item In cases when serverless applications are of high importance, the errors on cloud provider's side might silently cause inaccuracies in execution of the application code, e.g., double executions, which also have to be traced.
    In such cases, the sensitive parts of the application can be written as smart contract functions, because the fact whether they were executed or not cannot be influenced by the cloud provider and the inputs passed to them are immutably stored in the blockchain data structure.
    Furthermore, the blockchain protocol ensures that they are executed exactly the way they were designed.
    A smart contract used for this purpose can then pass the control back to a serverless component (using an event), or forward it to another smart contract function.
\end{compactenum}

\subsubsection{Pitfalls to Consider}
The first issue is that serverless components, such as FaaS functions, wishing to invoke smart contract functions need to consider specific requirements unique to blockchains.
In summary, one needs to ensure that the component
\begin{inparaenum}[(i)]
    \item supports a chosen language runtime and blockchain-specific client libraries for this language,
    \item has access to a blockchain node that facilitates the interaction with the blockchain system, and
    \item has a blockchain account that has enough funds/permissions to execute the designated smart contract function.
\end{inparaenum}
In \cref{sec:system-support}, we show how common providers approach these requirements.

Furthermore, smart contract function invocations can take several minutes to be confirmed in case the involved blockchain system utilizes the Proof-of-Work (PoW) consensus mechanism~\cite{Nakamoto2008}.
This means that serverless components, such as FaaS functions, cannot invoke them synchronously since this will likely violate the execution duration restrictions imposed by the platform provider.
Instead, the invocation should be asynchronous without waiting for confirmations (fire and forget).
Things get complicated when the request is of high value, e.g., it is handling a large sum of money.
In this case, the logic of the serverless application, as a whole, might require to check if the invocation succeeds.
To this end, special middleware components utilizing protocols, such as the Smart Contract Invocation Protocol (SCIP)~\cite{Falazi2020} can be used.
These components monitor the execution of smart contract functions, and inform the invokers about their success or failure.
Such components need to be hosted by the platform provider, or by a third-party.

Moreover, if the smart contract function is to be included in a sequence of event-driven invocations, i.e., it needs to further pass control to \enquote{the next} serverless function via emitting an event, this results in needing to tackle the same challenges we discussed in~\cref{sec:blockchains-as-event-emitters}.

On the other hand, smart contract functions are more restricted than typical FaaS functions.
For example, since blockchains process all requests sequentially and in full-replication, i.e., involving all of the system nodes, smart contract function executions are typically short, and process small amounts of data.
This is ensured by certain mechanisms, such as Ethereum's Gas~\cite{wood2018ethereum}.
Therefore, parallel data processing scenarios common for the FaaS world are not feasible to implement using smart contracts.
Furthermore, the direct costs of executing smart contracts on public blockchains are considerably higher than cloud-based alternatives.
For example, a study from 2017 showed that executing certain tasks on Ethereum costs 2 orders of magnitude more than their counterparts in AWS~\cite{Rimba2017}.
Considering recent pricing for the Ether cryptocurrency, the difference has now risen to about 3 orders of magnitude.
Thus, it is a trade-off between cost and trust that needs to be considered in each case individually.
Moreover, the developers of the serverless application must have the skills to represent the intended logic using smart contracts, which implies the need to learn a new programming language in certain cases, e.g., Solidity for Ethereum, and also implies the need to learn how to avoid a whole vector of security vulnerabilities specific to this domain~\cite{Sayeed2020SC_Attacks}.
Furthermore, supporting a DevOps experience similar to regular serverless applications is not a trivial task; first, the platform needs to support the automated deployment of the developed smart contracts to the underlying blockchain systems, and second, for some blockchains, the address of a smart contract is the digest of its bytecode.
This imposes challenges on release versioning, since new versions of a smart contracts will have different addresses than before, confusing client applications.
To tackle this issue, a specialized naming service can be used, e.g., the Ethereum Name Service~\cite{ethereum:ens} decouples the physical address of the smart contract from the address used to reach it.

\subsection{Scenario 3: Blockchains for Function Orchestration}
\label{sec:scenario-function-orchestration}
In this scenario, we analyze how blockchains and smart contracts can be used in function orchestration scenarios.

\subsubsection{Motivation and Analysis of Serverless Properties}
As discussed in~\Cref{sec:bg}, function orchestrators are a serverless component type, which acts as a basic business process management systems for coordinating the execution of function workflows.
Moreover, \cref{subsec:scenario-3} discusses cases in which smart contracts functions can be useful alternatives to FaaS functions.
It immediately follows that smart contract functions also could be incorporated into function orchestrations.
On the other hand, it has been shown that smart contracts can be used for implementing collaborative business processes or choreographies~\cite{DiCiccio2019_BcChoreographies} themselves.
For example, smart contracts can store messages exchanged between partners in an immutable and accountable fashion, or even implement the actual collaboration logic.
The major benefit is that such blockchain-based collaborations do not depend on third-parties any more, removing an obstacle that can, otherwise, hinder their effectiveness and performance altogether~\cite{Panayides2009}.

\subsubsection{Goal}
Considering the aforementioned similarities, the main focus of this scenario is to understand \textit{how to incorporate smart contract functions in function workflows, and how to employ the blockchain technology to enhance the trust guarantees provided by function orchestration scenarios}.

\begin{figure}[bp!]
    \begin{subfigure}[t]{\columnwidth}
        \centering
        \includegraphics[width=.95\columnwidth]{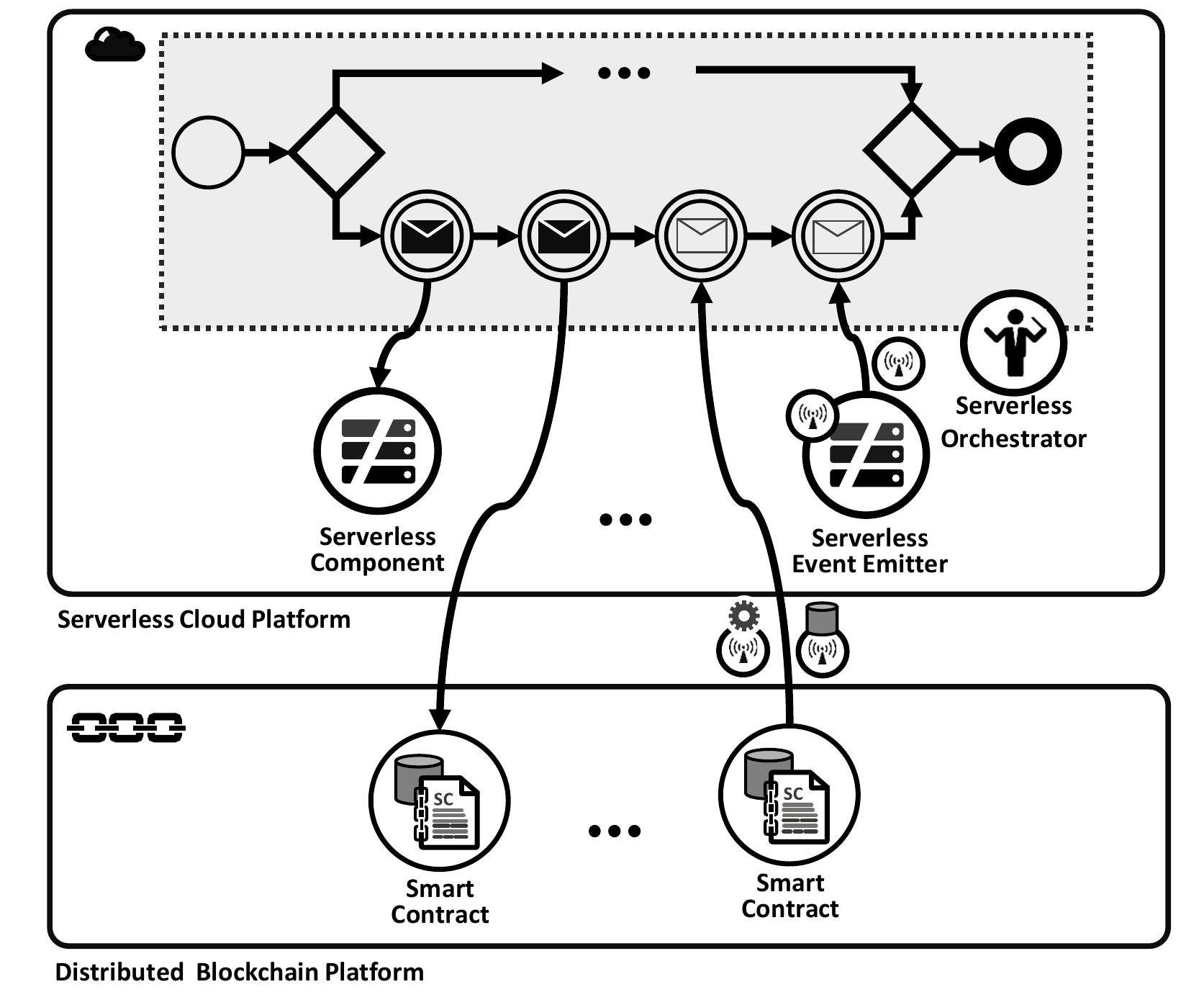}
        \caption{Smart contracts as building blocks in serverless workflows.}
        \label{fig:orchestration-a}
        \vspace{2em}
    \end{subfigure}    
    \begin{subfigure}[t]{\columnwidth}
        \centering
        \includegraphics[width=.95\columnwidth]{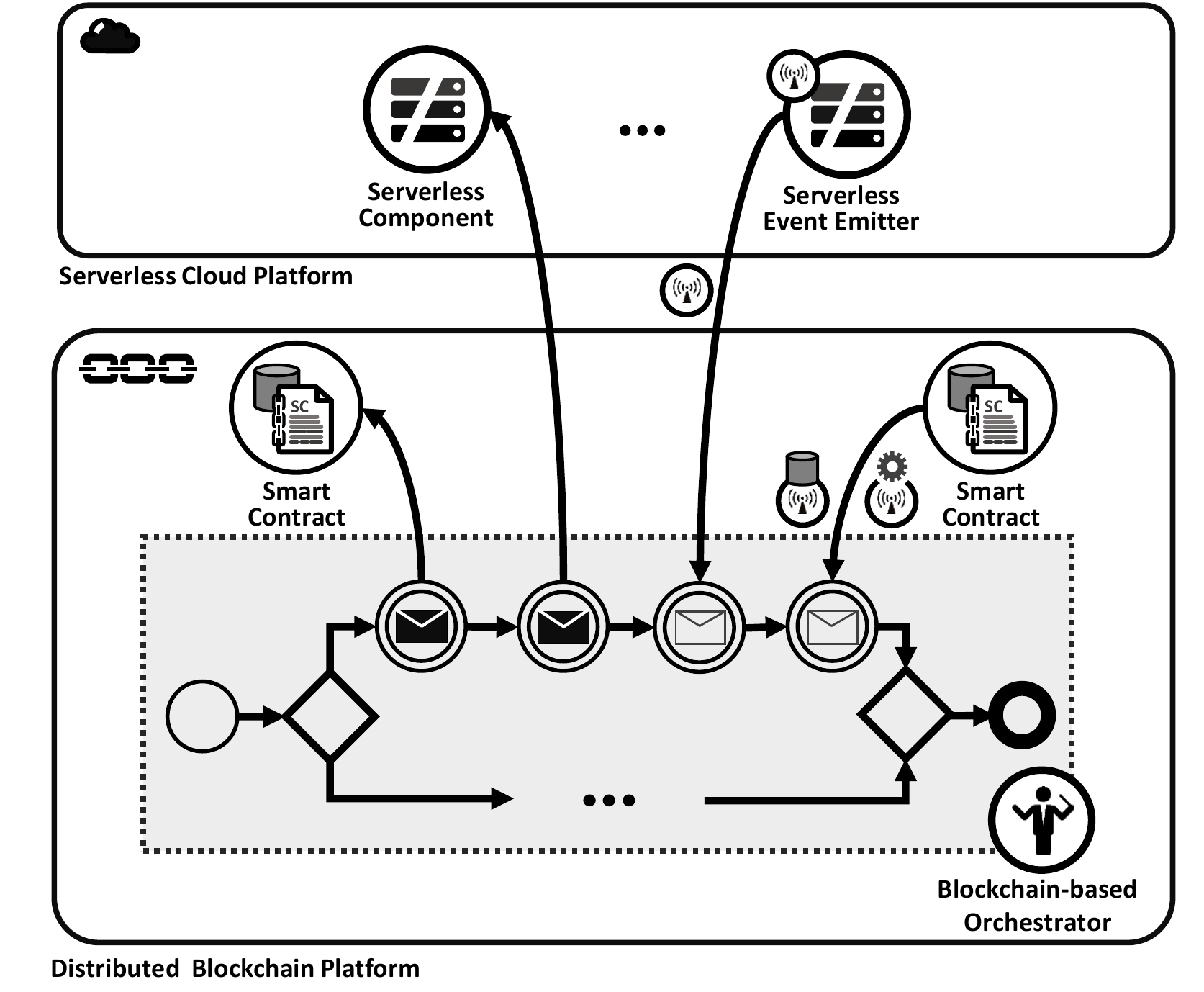}
        \caption{Blockchain-based workflow engines as serverless orchestrators with enhanced trust guarantees.}
        \label{fig:orchestration-b}
    \end{subfigure}
    \caption{Two ways of using blockchains in function orchestration.}
    \label{fig:orchestration}
\end{figure}

\subsubsection{Realization}
Essentially, blockchains and smart contracts can be used in function orchestration scenarios in two major ways, namely (i) for implementing specific parts of desired function workflows and (ii) as alternatives for function orchestrators themselves.
Firstly, as discussed in~\cref{subsec:scenario-3}, in certain contexts, smart contracts can be treated as FaaS functions.
Hence, this scenario can be relevant in the context of function orchestration, i.e., function workflows can combine regular- and smart contract function invocations.
Moreover, we have shown in~\cref{sec:blockchains-as-event-emitters} that smart contracts can act as event sources for serverless components.
This can also be applied here by allowing smart contracts to activate steps in function workflows via sending blockchain-specific events, as depicted in \cref{fig:orchestration-a}.

Secondly, employing a blockchain-based process engine, like Caterpillar~\cite{Lopez-Pintado2017}, as a serverless function orchestrator would store the state of all executed instances as well as the logic for advancing this state directly on the blockchain.
This achieves an immutable record of all state changes and ensures that the serverless platform provider cannot alter the process logic, which is very important in sensitive workflows.
Hence, a blockchain-based function orchestrator is provided to platform clients similar to a hosted service that can be used directly without managing it.

\subsubsection{Pitfalls to Consider}
The exciting properties of immutable and accountable storage of process state changes in a public blockchain come at a relatively high cost.
It is shown that the cost of executing similar process instances both via Ethereum smart contracts and Amazon Simple Workflow (SWF)~\cite{aws:swf} incurs more expensive costs in the former case up to 3 orders of magnitude considering the current exchange rate\footnote{1 ether = 300 USD (July 26, 2020). Source: \url{https://markets.businessinsider.com/currencies/eth-usd} } for the \textit{ether} cryptocurrency~\cite{Rimba2017}.
Furthermore, as mentioned earlier, data stored on-chain is limited in size. 
Thus, certain workflows with high data flow throughput are not suitable.
One solution for these problems is to perform computations or exchange data off-chain, while maintaining certain corresponding pieces of information on-chain to facilitate accountability~\cite{Kim2018}.
Moreover, in such scenarios the underlying function orchestrators must support handling smart contract invocations as activities in the workflow.
One way to facilitate such task for orchestrators that do not natively support integration with blockchains is to employ transformation-based approaches~\cite{falazi2019modeling,Falazi2019SC_Composition} applicable to the target function orchestrator's workflow definition style, e.g., a custom DSL or an orchestrating function.
When using such approaches, each activity representing a smart contract invocation is transformed into a set of standard constructs responsible for handling the invocation.

Finally, if a function orchestrator is hosted on the blockchain using smart contracts, the message exchange must happen in both directions, i.e., from serverless functions to the smart contracts and vice versa.
The former case entails the same technical difficulties discussed in~\cref{subsec:scenario-3}.
However, the latter case is especially problematic, since blockchains cannot actively access the external world~\cite{Al-Breiki2020}.
As we have discussed earlier, the reason behind this is that the execution of smart contracts needs to be deterministic, so they cannot involve sources of randomness, such as invocations to external systems.
Therefore, to realize a bidirectional message flow, systems like Caterpillar depend on off-chain components known as oracles~\cite{Al-Breiki2020}.
In general, if a smart contract function wants to invoke a service of an external system, it a emits a specific event that is captured by the oracle.
Then the oracle invokes the designated service on behalf of the smart contract and monitors its execution.
When it is done, the oracle delivers the result back to the corresponding smart contract by invoking a callback function.
A similar mechanism is used by Caterpillar to allow it to invoke external services, such as serverless components~\cite{Lopez-Pintado2017}.

\subsection{Scenario 4: Blockchains as Facilitators for the Integration of Serverless Applications}
\label{sec:integration}
In this scenario, we discuss how blockchains can act as a middleware that facilitates the integration of serverless applications in the same platform or on different platforms.

\subsubsection{Motivation and Analysis of Serverless Properties}
In the previous scenarios, blockchains and smart contracts were considered in the context of a single serverless application, i.e., when a certain application needs to utilize the blockchain technology to achieve certain desirable properties difficult to achieve otherwise.
However, another prominent use case of blockchains is to provide an accountable middleware to integrate applications and systems~\cite{Xu2016_BcSoftwareConnector}.
This can be very helpful when multiple serverless applications that belong to the same platform or to different platforms need to be integrated together, i.e., they need to exchange data or commands to achieve a common goal.
This task is generally difficult when maintaining trust is a requirement: in the case when applications are on the same platform, the provider needs to be trusted to deliver the integration messages exactly as intended by the relevant parties.
Things even get more difficult when the applications are on different platforms, since an additional trusted third-party is required to deliver the messages between them.
Therefore, a blockchain system can be used as a general-purpose software connector that provides certain services to facilitate the integration process, such as communication, coordination, and conversation services, while achieving desirable properties, like Byzantine fault tolerance, maintaining a total order of messages, and ensuring accountability~\cite{Xu2016_BcSoftwareConnector}.
Finally, this scenario can be considered as a generalization of all previously discussed scenarios, meaning that all characteristics related to them are relevant in this context as well.

\subsubsection{Goal}
The high-level goal of this scenario is to utilize \textit{blockchains and smart contracts to allow cross-application/cross-platform message exchange in a way that does not depend on trusted third-parties and ensures accountability in serverless architectures}.
As a result, messages crossing application or platform boundaries are attributable to their senders in a way that prevents them to repudiate this fact, and prevents forging their identities.

\begin{figure}[bp!]
    \centering
    \includegraphics[width=.95\columnwidth]{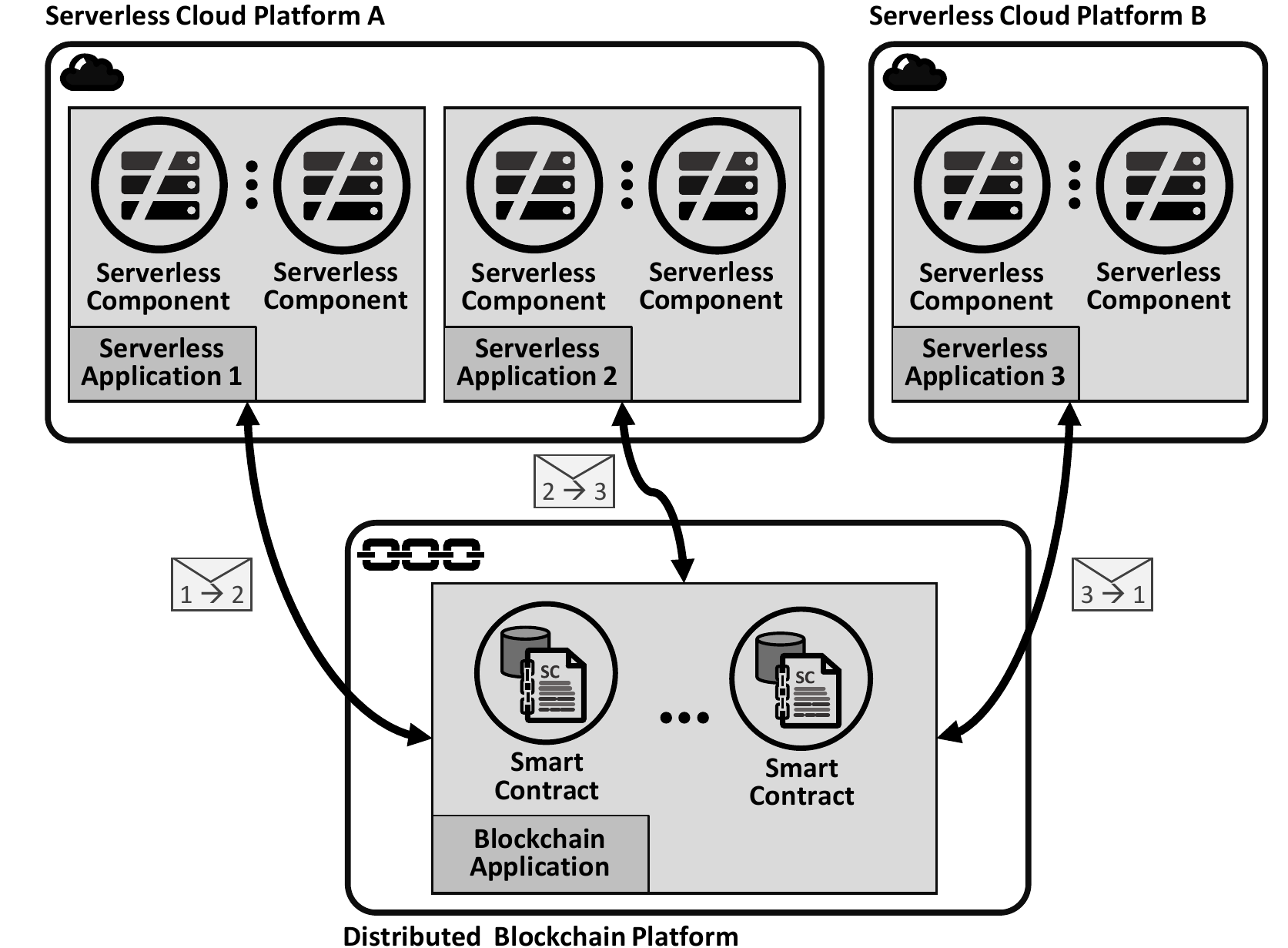}
    \caption{Smart contracts as accountable Message Bus for serverless platforms/applications.}
    \label{fig:sc-as-integrators}
\end{figure}

\subsubsection{Realization}

One possible way to define the role of blockchains in the integration problem for a set of distinct serverless cloud environments is to look at the previously discussed use cases through the prism of two well-known Enterprise Integration Patterns~(EIP) discussed by Hohpe and Woolf~\cite{hohpe2004enterprise}, namely the \textit{Message Bus} and \textit{Process Manager} patterns.
Firstly, a set of specially-tailored smart contracts can be used to implement a general cross-cloud event-driven execution, with these smart contract serving abstractly as an accountable implementation of the Message Bus pattern.
We refer to this serverless integration scenario as \textit{Blockchain as Message Bus} since the overall idea is applicable not only to serverless contexts.
This is done by a set of smart contracts designed to receive request messages, i.e., blockchain transactions, from source components, and as a result emit smart contract events that contain, e.g., the sender information, the payload of the original message, and the information required to identify the target component, which can then detect this event and consume the associated message (see \cref{fig:sc-as-integrators}).
As in the Message Bus pattern, the to-be-integrated serverless environments share (i) a common data model~(blockchain "integration" events), (ii) a common set of commands~(invocation of smart contracts), and (iii) an infrastructure~(the blockchain itself) that actually enables communication among involved applications via the same set of interfaces.
The main benefit of using the blockchain as a message bus is to immutably store exchanged messages and to ensure that the routing logic is executed exactly as designed.
Another advantage is the guarantee that the smart contract-based message bus is always available due to the very high replication degree of the underlying (public) blockchain.

It is worth highlighting that the \textit{Blockchain as Message Bus} integration scenario relies on a combination of two use cases discussed previously, namely (i) \textit{Utilizing Blockchains as Event Emitters} discussed in~\Cref{sec:blockchains-as-event-emitters}, and (ii) \textit{Utilizing Smart Contracts as FaaS Functions} discussed in~\Cref{subsec:scenario-3}.
Event emitting is required to notify one or more subscribed serverless environments about the integration event, whereas the invocation of smart contract functions is necessary to trigger the execution of the integration logic embedded in them by the serverless component initiating the integration flow.
An obvious downside of this scenario is that the integration logic is hard-coded into smart contracts, which complicates making changes.

As was discussed in~\Cref{sec:scenario-function-orchestration}, blockchains can also be used as workflow engines, e.g., Caterpillar~\cite{Lopez-Pintado2017}, which enables the \textit{Blockchain as Process Manager} integration scenario.
In contrast with the Blockchain as Message Bus integration scenario, the interaction logic is not hard-coded into a predefined set of smart contracts, but rather modeled using a common business process language, such as BPMN, and then deployed to the blockchain-based process engine, which is responsible for transforming it into a set of model-specific smart contracts that enact the actual integration workflow.
The major benefits over the first case include decoupling the integration logic from the middleware executing it, and the usage of well-known languages to model it.
The overall interaction is similar to the one shown in~\Cref{fig:orchestration-b}, but with serverless environments interacting via the blockchain-enacted workflow.

\subsubsection{Pitfalls to Consider}
Since both \textit{Blockchain as Message Bus} and \textit{Blockchain as Process Manager} integration options are rather combinations of scenarios discussed in~\Cref{sec:blockchains-as-event-emitters,subsec:scenario-3,sec:scenario-function-orchestration}, they inherit the already discussed challenges from each corresponding scenario.
For example, the \textit{Blockchain as Message Bus} has the same challenges discussed in~\Cref{sec:blockchains-as-event-emitters,subsec:scenario-3}.
Moreover, the challenges described in~\Cref{sec:scenario-function-orchestration} apply as well in case function worklfows are being integrated using blockchains as a message bus. 
This is also true for the \textit{Blockchain as Process Manager} option, since it combines all use cases together, i.e., it incorporates emitting and processing blockchain events, smart contracts carrying the integration logic, as well as the orchestration-related challenges discussed in~\Cref{sec:scenario-function-orchestration}.

Another difficulty, which is related to the topic of integration in general, is that even without needing to trust a third-party, integration scenarios require participants to agree on an overall joint process.
Moreover, the deployment automation for such composite application topologies, i.e., that combine serverless deployments with blockchain-specific deployments, is a non-trivial question to answer.
As we will see in \cref{sec:system-support}, only few platform providers offer solutions to this problem.

\section{Technical Requirements For Using Blockchains in Serverless Architectures}
\label{sec:requirements}
In this section, based on the realization details and pitfalls discussed in the scenarios of~\cref{sec:scenarios}, we formulate the requirements that must be fulfilled by the target hosting platform to enable the usage of blockchains and smart contracts in serverless architectures.
These requirements are categorized into three larger groups which are discussed in-detail in the following.
Moreover, after discussing the requirements, we elaborate how the aforementioned requirements influence the scenarios discussed in~\Cref{sec:scenarios} as shown in~\Cref{tab:requirement-scenario-mapping}.

\subsection{Blockchain Node Access Requirements}
\label{req:group-A}
This group encompasses requirements related to accessing a blockchain node, i.e., serverless components willing to interact with the blockchain must be able to issue requests to a blockchain node and receive the responses back.

\subsubsection{Blockchain Node is a Serverless Component}
\label{req:A-1}

\begin{center}
    \vspace{.5\baselineskip}
    \noindent
    \setlength{\fboxsep}{2mm}
    \fbox{
        \begin{minipage}{.9\linewidth}
            \textit{\textbf{Req.~A-1:} The hosting type of the blockchain node which is used in a serverless architecture for interacting with a blockchain system must comply with the serverless architecture style.}
    \end{minipage}}
    \setlength{\fboxsep}{2mm}
    \vspace{.2\baselineskip}
\end{center}

\noindent
The first requirement is that if the blockchain is planned to be used as a component in serverless architectures, as it is relevant for all scenarios discussed in~\cref{sec:scenarios}, it has to be used in a serverless manner.
This means that a blockchain node, which is required for any kind of interaction with the underlying blockchain system, has to be managed by a third party eliminating the need to host and manage the node or the entire network of nodes on premises.

\subsubsection{Access to Authorized Blockchain Account}
\label{req:A-2}

\begin{center}
    \vspace{.5\baselineskip}
    \noindent
    \setlength{\fboxsep}{2mm}
    \fbox{
        \begin{minipage}{.9\linewidth}
            \textit{\textbf{Req.~A-2:} Serverless applications employing the blockchain technology must have access to a blockchain account with sufficient permissions.}
    \end{minipage}}
    \setlength{\fboxsep}{2mm}
    \vspace{.2\baselineskip}
\end{center}

\noindent
Interactions with blockchain systems are authenticated, meaning that a blockchain account must be used by client applications when issuing requests to the network.
This usually means that they must be digitally signed by a private key, and that the network peers are able to verify this signature via some sort of a public certificate.
Furthermore, the question of having proper permissions associated with the blockchain account has to be addressed.
This does not only apply to permissioned blockchains, but also to permissionless blockchains if the invoked smart contract has access control implemented in some way, e.g., via Solidity Function Modifiers~\cite{ethereum:functionModifiers}, which facilitate restricting the transactions allowed to interact with a given smart contract function by specifying the blockchain identities allowed to be senders or originators of these transactions.
Therefore, serverless applications must have a blockchain account with proper permissions and have to be able to formulate blockchain transactions and sign them using the account's credentials either directly or via third-party services performing these tasks on behalf of it.

This requirement affects all scenarios dicsussed in~\cref{sec:scenarios}. 
Firstly, it is always relevant when invoking smart contract functions, hence affecting Scenarios~2-4.
Additionally, it is also relevant when monitoring blockchain events for certain blockchains that require the client to be authenticated and authorized even for read-only operations, e.g., Hyperledger Fabric~\cite{androulaki2018fabric}, hence affecting Scenario~1 in some cases.

\subsubsection{Sufficient Funds for Transaction Processing}
\label{req:A-3}

\begin{center}
    \vspace{.5\baselineskip}
    \noindent
    \setlength{\fboxsep}{2mm}
    \fbox{
        \begin{minipage}{.9\linewidth}
            \textit{\textbf{Req.~A-3:} Serverless applications employing the blockchain technology might need to account for blockchain transaction fees.}
    \end{minipage}}
    \setlength{\fboxsep}{2mm}
    \vspace{.2\baselineskip}
\end{center}

\noindent
Certain blockchain systems, especially permissionless ones, require client applications to pay fees for each submitted transaction. 
As such, the cost incurred by serverless application must also include the fees for state-changing communication with blockchains of these types (this applies to all scenarios that involve smart contract invocations, i.e., all scenarios discussed in \cref{subsec:scenario-3,sec:scenario-function-orchestration,sec:integration}).
For example, applications can be charged with these fees directly, or via third-party services depending on how the blockchain account is managed (\req{A-2},~\Cref{req:A-2}).

\subsubsection{Access to External Blockchain Nodes}
\label{req:A-4}

\begin{center}
    \vspace{.5\baselineskip}
    \noindent
    \setlength{\fboxsep}{2mm}
    \fbox{
        \begin{minipage}{.9\linewidth}
            \textit{\textbf{Req.~A-4:} Serverless applications employing the blockchain technology might require direct or indirect interaction with external blockchain nodes.}
    \end{minipage}}
    \setlength{\fboxsep}{2mm}
    \vspace{.2\baselineskip}
\end{center}

\noindent
To support deployments in which not the whole blockchain network is hosted by a single cloud platform, e.g., in the case of multi-cloud or hybrid deployments of permissioned blockchains, or in the case of permissionless blockchains, the blockchain node that the serverless application communicates with (\req{A-1},~\Cref{req:A-1}) must be able to interact with external nodes, which, for example, might not be possible when using specific Blockchain-as-a-Service (BaaS) offerings.
This requirement is relevant for all discussed scenarios when using public blockchains, or if trust in a single provider is to be avoided, which is especially important in cross-platform integrations (see \cref{sec:integration}).

\subsection{Inter-component Interaction Requirements}
\label{req:group-B}
This group of requirements focuses on the interaction between traditional serverless application components and blockchains, e.g., subscription to blockchain events, invocation of smart contracts, etc.
Clearly, all of these requirements also demand that access to a blockchain node is provided (see requirements of \cref{req:group-A}).

\subsubsection{Support for Blockchain Events Subscription}
\label{req:B-1}

\begin{center}
    \vspace{.5\baselineskip}
    \noindent
    \setlength{\fboxsep}{2mm}
    \fbox{
        \begin{minipage}{.9\linewidth}
            \textit{\textbf{Req.~B-1:} Serverless applications employing the blockchain technology might need to be able to subscribe to blockchain events.}
    \end{minipage}}
    \setlength{\fboxsep}{2mm}
    \vspace{.2\baselineskip}
\end{center}

\noindent
As we have seen in all discussed scenarios, serverless applications might need to be able to subscribe to different types of events originating from blockchain systems.
However, one caveat in using such events in serverless architectures is that blockchains are not meant to be aware of external systems.
As a result, the integration scenarios in which events emitted from blockchain components, e.g., the consensus component or a smart contract are \textit{synchronously pushed} to the desired target, become impossible.
This leads to the requirement \textit{to poll for blockchain events} either directly by a combination of serverless components that belong to the target application, or via a third-party managed service.
This requirement is obviously relevant for all the scenarios that require listening to blockchain events (see \cref{sec:blockchains-as-event-emitters,sec:scenario-function-orchestration,sec:integration}), and it is also relevant in case a smart contract function acting as a servelress function needs to emit events to be captured by other serverless components (see \cref{subsec:scenario-3}).

\subsubsection{Support for Smart Contract Invocation}
\label{req:B-2}

\begin{center}
    \vspace{.5\baselineskip}
    \noindent
    \setlength{\fboxsep}{2mm}
    \fbox{
        \begin{minipage}{.9\linewidth}
            \textit{\textbf{Req.~B-2:} Serverless applications employing the blockchain technology might need to be able to invoke smart contracts.}
    \end{minipage}}
    \setlength{\fboxsep}{2mm}
    \vspace{.2\baselineskip}
\end{center}

\noindent
Implementing the invocation of smart contracts from a serverless component has some pitfalls too.
Firstly, in addition to having access to a third-party managed blockchain node (\req{A-1},~\Cref{req:A-1}), the serverless application also needs to have sufficient permissions and funds for the invocation (\req{A-2} and \req{A-3} in~\Cref{req:A-2,req:A-3}).
Furthermore, in case the invocation originates from a FaaS-hosted function, the required language runtime and software libraries must be supported by the chosen FaaS platform.
The same applies to function workflows: if a function workflow is to contain smart contract invocation steps, they have to be modeled within it, which is a task that highly depends on the target orchestrator.
Some orchestrators require the workflow to be modeled using a custom DSL, whereas others rely on orchestrating functions.
These formats have to support smart contract invocations.
Moreover, the invocation message format, i.e., the blockchain transaction, and the way the target smart contract is addressed are blockchain-specific which means that the component sending the invocation request to the blockchain node must use the exact format and addressing scheme expected by the chosen blockchain.
Obviously, this requirement is relevant for all scenarios that involve the invocation of smart contract functions (see scenarios described in~\cref{subsec:scenario-3,sec:scenario-function-orchestration,sec:integration}).

\subsubsection{Transaction Durability Guarantees}
\label{req:B-3}

\begin{center}
    \vspace{.5\baselineskip}
    \noindent
    \setlength{\fboxsep}{2mm}
    \fbox{
        \begin{minipage}{.9\linewidth}
            \textit{\textbf{Req.~B-3:} Serverless applications employing the blockchain technology might need to be able to ensure transactions' durability.}
    \end{minipage}}
    \setlength{\fboxsep}{2mm}
    \vspace{.2\baselineskip}
\end{center}

\noindent
Depending on the consensus algorithm employed by the blockchain, it might be necessary to wait for a specific amount of time after a submitted blockchain transaction was confirmed to ensure its durability.
This might be problematic when, e.g., the transaction represents a smart contract invocation originating from a FaaS function since not all platforms tolerate long-running tasks.
Therefore, in cases when the transaction durability is important, special application design measures and potentially supporting third-party services might need to be considered.
This requirement is relevant for all scenarios that involve invocation of smart contract functions (see \cref{subsec:scenario-3,sec:scenario-function-orchestration,sec:integration}).

\subsubsection{Blockchain as Active Communicator}
\label{req:B-4}

\begin{center}
    \vspace{.5\baselineskip}
    \noindent
    \setlength{\fboxsep}{2mm}
    \fbox{
        \begin{minipage}{.9\linewidth}
            \textit{\textbf{Req.~B-4:} The business logic of certain serverless applications might require the underlying blockchain system to start interaction with external components.}
    \end{minipage}}
    \setlength{\fboxsep}{2mm}
    \vspace{.2\baselineskip}
\end{center}

\noindent
While blockchains were not originally intended to interact with the outside world actively, i.e., by sending requests, there are mechanisms allowing such interaction without forcing the external component to be aware of the specifics of blockchain systems.
Blockchain-external components, known as \textit{oracles}~\cite{Al-Breiki2020}, might be used to serve as intermediaries that decouple the blockchain from the outside systems by subscribing to specific blockchain events, and actively invoking the designated system on behalf of the blockchain.
This results in relaxing the system's decentralization level, but might be a requirement in certain cases.
Specifically, in \textit{Scenario 3: Blockchains for Function Orchestration} (see \cref{sec:scenario-function-orchestration}), we saw that the application logic can be stored and managed by a blockchain-based orchestrator.
If the logic involves invocations to external systems, oracles will be required as discussed above.
This is also relevant when integrating multiple serverless platforms using blockchains (see \cref{sec:integration}) if the target platform provides no native integration support.

\subsection{DevOps Requirements}
\label{req:group-C}
This group of requirements focuses on aspects related to development and operations of serverless applications that employ blockchains as application components.

\subsubsection{Support for Smart Contracts Development}
\label{req:C-1}

\begin{center}
    \vspace{.5\baselineskip}
    \noindent
    \setlength{\fboxsep}{2mm}
    \fbox{
        \begin{minipage}{.9\linewidth}
            \textit{\textbf{Req.~C-1:} Specialized tools might be needed to allow developers to properly develop smart contracts.}            	
    \end{minipage}}
    \setlength{\fboxsep}{2mm}
    \vspace{.2\baselineskip}
\end{center}

\noindent
As discussed in \textit{Scenario 2: Smart Contracts as Serverless Functions} (see~\cref{subsec:scenario-3}), the requirements for implementing smart contracts highly depend on the chosen blockchain vendor.
Firstly, the expertise in system-specific restrictions is required, e.g., specialized programming languages, data structures and persistence mechanisms, quotas on inputs and outputs.   
In addition, implementation of smart contracts must keep security (anti-)patterns in mind.
Another important part is understanding which business logic is feasible to place in a smart contract function since multiple aspects need to be considered, e.g., read-only vs. read-write functions, data persistence, data throughput, computational efficiency, etc.
Finally, debugging smart contracts is not a trivial task due to the distributed nature of the hosting blockchains.
Therefore, it is beneficial if the platform provides sufficient tooling, e.g., IDEs or IDE extensions for smart contract development, smart contract templates for common cases, simulation and debugging tools, code analysis tools for detecting anti-patterns, etc., to help developers efficiently create or edit smart contracts with better quality.
Obviously, this is a non-functional requirement relevant only to scenarios in which serverless application developers need to deal with smart contract development (see \cref{subsec:scenario-3}).

\setlength\tabcolsep{4pt}
\begin{table*}[bp]
    \centering
    \begin{threeparttable}[b]
        \centering
        \captionsetup{justification=raggedright,singlelinecheck=false,format=hang}
        \caption{The effect of identified requirements for using blockchains in serverless architectures on the scenarios discussed in \cref{sec:scenarios}. We use the \ding{51} symbol if a requirement affects a scenario, with additional comments provided as a note (if applicable). }
        \label{tab:requirement-scenario-mapping}
        \scriptsize
        \def\arraystretch{1.5}
        \definecolor{lightgray}{gray}{0.9}
        \begin{tabularx}{.95\textwidth}
            {@{} 
                c 
                m{35.3mm} @{}
                m{23mm}     
                m{22.6mm}     
                m{17mm}     
                m{17mm}     
                m{2mm}  @{}    
                m{17mm}     
                m{17mm}     
            }
            \hline
            \multicolumn{2}{l}{}
            & \multicolumn{7}{c}{\textit{Affected Scenarios}} \\
            \cline{3-9}
            
            \multicolumn{2}{l}{}
            & \multirow{2}{21mm}{\centering Scenario 1: Blockchains as Serverless Event Emitters (\Cref{sec:blockchains-as-event-emitters})}
            & \multirow{2}{21mm}{\centering Scenario 2: Smart Contracts as Serverless Functions (\Cref{subsec:scenario-3})}
            & \multicolumn{2}{m{32mm}}{\centering Scenario 3: Blockchains for Function Orchestration (\Cref{sec:scenario-function-orchestration})}
            &    
            & \multicolumn{2}{m{34mm}}{\centering Scenario 4: Blockchains as Facilitators For the Integration of Serverless Applications(\Cref{sec:integration})} \\ \cline{5-6} \cline{8-9}

            \multirow{3}{*}{}
            & \multirow{3}{*}{}
            & \multirow{2}{*}{}
            & \multirow{2}{*}{}
            & \centering Orchestration a)  
            & \centering Orchestration b)  
            &    
            & \centering Message Bus      
            & \centering \arraybackslash Process Manager \\  
            \hline
            
            \parbox[t]{5mm}{\multirow{4}{*}[-3em]{\rotatebox[origin=c]{90}{\begin{minipage}{5em}\centering \textit{Req. Group A}\end{minipage}}}}
            
            & \cellcolor{lightgray} \centering Blockchain Node is a Serverless Component (\req{A-1})
            & \cellcolor{lightgray} \centering \ding{51} 
            & \cellcolor{lightgray} \centering \ding{51} 
            & \cellcolor{lightgray} \centering \ding{51} 
            & \cellcolor{lightgray} \centering \ding{51} 
            & \cellcolor{lightgray}   
            & \cellcolor{lightgray} \centering \ding{51}  
            & \cellcolor{lightgray} \centering \arraybackslash \ding{51} \\ 
            
            & \centering Access to Authorized Blockchain Account (\req{A-2})
            & \centering \ding{51}\tnote{(1)} 
            & \centering \ding{51} 
            & \centering \ding{51}\tnote{(2)} 
            & \centering \ding{51} 
            &    
            & \centering \ding{51}  
            & \centering \arraybackslash \ding{51}\tnote{(2)} \\  
            
            & \cellcolor{lightgray} \centering Sufficient Funds for Transaction Processing (\req{A-3})
            & \cellcolor{lightgray} \centering  
            & \cellcolor{lightgray} \centering \ding{51}\tnote{(3)} 
            & \cellcolor{lightgray} \centering \ding{51}\tnote{(3)} 
            & \cellcolor{lightgray} \centering \ding{51}\tnote{(2)}\tnote{(3)} 
            & \cellcolor{lightgray}   
            & \cellcolor{lightgray} \centering \ding{51}\tnote{(3)}  
            & \cellcolor{lightgray} \centering \arraybackslash \ding{51}\tnote{(2)}\tnote{(3)} \\  
            
            & \centering Access to External Blockchain Nodes (\req{A-4})
            & \centering \ding{51}\tnote{(4)} 
            & \centering \ding{51}\tnote{(4)} 
            & \centering \ding{51}\tnote{(4)} 
            & \centering \ding{51}\tnote{(4)} 
            &    
            & \centering \ding{51}  
            & \centering \arraybackslash \ding{51} \\ 
            \hline            
            
            \parbox[t]{5mm}{\multirow{4}{*}[-1.2em]{\rotatebox[origin=c]{90}{\begin{minipage}{5em}\centering \textit{Req. Group B}\end{minipage}}}}
            
            & \cellcolor{lightgray} \centering Support for Blockchain Events Subscription (\req{B-1})
            & \cellcolor{lightgray} \centering \ding{51} 
            & \cellcolor{lightgray} \centering \ding{51}\tnote{(5)} 
            & \cellcolor{lightgray} \centering \ding{51} 
            & \cellcolor{lightgray} \centering \ding{51} 
            & \cellcolor{lightgray}   
            & \cellcolor{lightgray} \centering \ding{51}  
            & \cellcolor{lightgray} \centering \arraybackslash \ding{51} \\  
            
            & \centering Support for Smart Contract Invocation (\req{B-2})
            & \centering  
            & \centering \ding{51} 
            & \centering \ding{51} 
            & \centering \ding{51} 
            &    
            & \centering \ding{51} 
            & \centering \arraybackslash \ding{51} \\  
            
            & \cellcolor{lightgray} \centering Transaction Durability Guarantees (\req{B-3})
            & \cellcolor{lightgray} \centering  
            & \cellcolor{lightgray} \centering \ding{51}\tnote{(3)} 
            & \cellcolor{lightgray} \centering \ding{51}\tnote{(3)} 
            & \cellcolor{lightgray} \centering \ding{51}\tnote{(3)} 
            & \cellcolor{lightgray}   
            & \cellcolor{lightgray} \centering \ding{51}\tnote{(3)} 
            & \cellcolor{lightgray} \centering \arraybackslash \ding{51}\tnote{(3)} \\  
            
            & \centering Blockchain as Active Communicator (\req{B-4})
            & \centering  
            & \centering  
            & \centering  
            & \centering \ding{51} 
            &    
            & \centering  
            & \centering \arraybackslash \ding{51}\tnote{(6)} \\  
            \hline
            
            \parbox[t]{5mm}{\multirow{2}{*}[.5em]{\rotatebox[origin=c]{90}{\begin{minipage}{5em}\centering \textit{Req. Group C}\end{minipage}}}}
            
            & \cellcolor{lightgray} \centering Support for Smart Contracts Development (\req{C-1})
            & \cellcolor{lightgray} \centering  
            & \cellcolor{lightgray} \centering \ding{51} 
            & \cellcolor{lightgray} \centering  
            & \cellcolor{lightgray} \centering  
            & \cellcolor{lightgray}   
            & \cellcolor{lightgray} \centering  
            & \cellcolor{lightgray} \centering \arraybackslash  \\  
            
            & \centering Support for Deployment Automation (\req{C-2})
            & \centering \ding{51} 
            & \centering \ding{51} 
            & \centering  
            & \centering \ding{51} 
            &    
            & \centering  
            & \centering \arraybackslash \\  
            \hline
            
        \end{tabularx}
        \begin{tablenotes}[normal,para]
            \scriptsize
            \item[(1)] \textit{Only for certain blockchains.}
            \item[(2)] \textit{Depends on the implementation of blockchain-based process engine.}
            \item[(3)] \textit{Only for blockchains with native cryptocurrency.}
            \item[(4)] \textit{Non-functional requirement (enhances trust guarantees).}
            \item[(5)] \textit{Only if smart contracts pass control back via events.}
            \item[(6)] \textit{Only if the target platform does not support native integration.}
        \end{tablenotes}
    \end{threeparttable}
\end{table*}

\subsubsection{Support for Deployment Automation}
\label{req:C-2}

\begin{center}
    \vspace{.5\baselineskip}
    \noindent
    \setlength{\fboxsep}{2mm}
    \fbox{
        \begin{minipage}{.9\linewidth}
            \textit{\textbf{Req.~C-2:} Serverless applications employing blockchain might need to be deployed automatically.}
    \end{minipage}}
    \setlength{\fboxsep}{2mm}
    \vspace{.2\baselineskip}
\end{center}

\noindent
While there are multiple deployment automation technologies for serverless applications, e.g., AWS SAM~\cite{aws:sam}, Serverless Framework~\cite{serverless:serverlessFramework}, or Azure Resource Manager~\cite{ms:azure-resourceManager}, at least three novel types of deployment tasks have to be automated with respect to blockchain-hosted parts of a serverless application, namely (i) the deployment of smart contracts, (ii) the deployment of blockchain-aware function workflows, and (iii) the creation of event bindings.
The deployment of new smart contracts is necessary in case one or more smart contract functions are to be treated as serverless functions (see \cref{subsec:scenario-3}).
Furthermore, it is also necessary when a blockchain-based process engine (like Caterpillar~\cite{Lopez-Pintado2017}) is used as a serverless function orchestrator since  workflows are transformed into model-specific smart contracts before being deployed to the underlying blockchain system (see \cref{sec:scenario-function-orchestration}).
Moreover, as we have seen in \textit{Scenario 3: Blockchains for Function Orchestration} in~\cref{sec:scenario-function-orchestration}, serverless function orchestrators might need to invoke smart contract functions during the execution of blockchain-aware workflows (\req{B-2},~\Cref{req:B-2}).
The association between a workflow step and a smart contract invocation must be handled if the workflow is to be automatically deployed.
On the other hand, two types of event bindings can be recognized in the discussed scenarios:
firstly, as we saw in \cref{req:B-1}, events emitted from blockchain systems sometimes need to trigger other serverless components (see \textit{Scenario 1: Blockchains as Serverless Event Emitters} in~\cref{sec:blockchains-as-event-emitters}), and secondly, serverless events might need to be configured to trigger the execution of a smart contract function (this is especially relevant for the scenario discussed in \cref{subsec:scenario-3}).


\section{Survey of Technologies Enabling Blockchains in Serverless Architectures}
\label{sec:system-support}
In this section, we investigate what is needed to realize the scenarios proposed in~\Cref{sec:scenarios} in the light of existing technologies. 
Therefore, we analyze the readiness of current serverless offerings to enable utilizing blockchains and smart contracts in serverless architectures, i.e., to what extent they fulfill the requirements identified in \cref{sec:requirements}.

Since blockchains can also be provided \enquote{as-a-service}, we define the concept of Blockchain-as-a-Service (BaaS) and its variations before introducing the survey itself.

\subsection{Blockchain-as-a-Service (BaaS)}
\label{sec:baas}
Blockchain-as-a-Service~(BaaS) is a category of cloud service offerings that enable utilizing the blockchain technology while minimizing the required management and maintenance tasks~\cite{Lu2019uBaaS}.
This promotes more focus on application development, both in the case of on-chain applications, i.e., smart contracts, and off-chain applications that utilize the blockchain, i.e., DApps.
BaaS is particularly helpful, since it outsources the skills and expertise required to deploy and manage blockchain networks, which is very beneficial, since the blockchain is a relatively new technology and the number of IT professionals capable of managing its infrastructure is limited.
On the other hand, available BaaS offerings can be categorized under various service models~\cite{Singh2018BaaSTrust}.
Notably, most BaaS offerings correspond to the PaaS service model in which the tenants themselves design, configure and instantiate entire blockchain networks based on their needs.
The platform, in this case, provides the necessary pre-built executables or images, and seamlessly manages tasks like node restarts, inter-node communication, and integration with other services.
Moreover, the platform usually provides some sort of an API that allows to perform managerial and operational tasks on the hosted blockchain instances.
Other BaaS offerings correspond to the Software-as-a-Service (SaaS) service model, whereby one or more providers run a special-purpose blockchain instance, and allow tenants, mostly enterprises, to interact with it using a high-level API.
Overall, due to the nature of serverless applications, BaaS offerings are good candidates for enabling the integration of blockchains into serverless architectures.

\subsection{Survey Design}
One of the entry-level requirements for implementing the scenarios described in~\Cref{sec:scenarios} is the possibility to access a blockchain node in a serverless fashion~(see \req{A-1}).
Although there exist third-party BaaS solutions not from major cloud providers, to avoid shifting the discussion mainly towards multi-cloud integration challenges, we restricted the survey to analyze the capabilities of only those cloud infrastructures that offer both BaaS and FaaS offerings due to the decisive role FaaS plays in our scenarios.
Hence, we defined the following inclusion criteria for selecting BaaS offerings intended for analysis:
\begin{itemize}
    \item[\ding{51}] It must belong to a public cloud provider, which has at least one general-purpose FaaS platform.
    \item[\ding{51}] Its documentation must be available in English.
\end{itemize}

Since the presence of a FaaS offering in a public cloud was required by our selection criteria, we referred to a list of FaaS offerings continually maintained by the Cloud Native Computing Foundation (CNCF)~\cite{LinuxFoundation2020CNCF} as the main data source for identifying suitable BaaS offerings.
CNCF is part of the Linux Foundation and it generally aims at promoting cloud native technologies, also including the list of up-to date serverless technologies, e.g., FaaS platforms and FaaS-specific deployment automation tools.
At the moment of writing~(October 2020), the aforementioned list contains a set of 31 FaaS platforms including hosted and installable solutions.
After applying the inclusion criteria, only 7/31 entries from the list had both FaaS and blockchain-specific service offerings.
This list comprised the following public cloud providers: \textit{(1)~Alibaba Cloud, (2)~Amazon Web Services, (3)~Cloudflare, (4)~Huawei Cloud, (5)~IBM Cloud, (6)~Microsoft Azure, and (7)~Oracle Cloud.}
As a next step, we analyzed the documentation of each provider's BaaS offering individually, focusing on the degree to which the requirements discussed in \cref{sec:requirements} are fulfilled, and as a result identifying which scenarios are applicable and to what level (see \cref{sec:scenarios}).
Finally, we compared the analyzed cases and clarified the overall picture of blockchain integration into serverless architectures as depicted by the current technology landscape.

\subsection{Analysis of Public Cloud BaaS Offerings}
Next, we individually analyze each of the 7 considered FaaS platform providers (ordered alphabetically).

\subsubsection{Alibaba Cloud}
Alibaba Cloud BaaS~\cite{alibaba:baas} is a comprehensive offering that currently supports three permissioned blockchains: (i) Hyperledger Fabric~\cite{androulaki2018fabric} (ii) ConsenSys Quorum~\cite{consensys:quorum}, and (iii)~Ant Blockchain~\cite{antgroup:antchain}.
The offering provides an intuitive console that allows deploying new instances of these blockchains on Kubernetes clusters provisioned by the Alibaba ACK~\cite{alibaba:ack} service (see \req{A-1}). 
Furthermore, the console facilitates monitoring and managing already deployed instances, and allows other enterprises to join them to form consortia.
At the time-being, only public and private cloud deployments are possible, and blockchain instances cannot include external nodes, e.g., multi-cloud or hybrid deployments.
An exception is ConsenSys Quorum, which allows adding external validator nodes to the managed blockchain instance (see \req{A-4}).
Moreover, Alibaba Cloud BaaS allows the deployment and upgrade of smart contracts, and has a tool to analyze their code and provide warnings on potential vulnerabilities.
Besides, an integrated IDE, called Cloud IDE, allows the creation and debugging of Ant Blockchain smart contracts (see \req{C-1} and \req{C-2}).
Additionally, Alibaba Cloud BaaS can be integrated with other cloud services.
Notably, the platform can be configured to allow blockchain smart contract events to trigger the execution of serverless functions hosted on the Function Compute~\cite{alibaba:functioncompute} service (see \req{B-1}).
They can also be published on RocketMQ~\cite{alibaba:rocketmq}, Alibaba's managed message queue service, or be sent to an external HTTP API (see \req{B-4}).
Finally, a REST API provided on top of the BaaS offering allows the execution of smart contract functions, in addition to querying past events and transactions, and subscribing to future ones (see \req{B-2}).
This API can be utilized from the Function Compute service, which eases the integration with serverless applications.
To ensure that the REST API is only used by authorized users, tokens generated by the Resource Access Management (RAM)~\cite{alibaba:ram} service must accompany request messages.
RAM provides a mechanism to associate cloud users with blockchain accounts and manage their permissions (see \req{A-2}).

Considering the set of capabilities provided by Alibaba Cloud BaaS offering, the platform can achieve only two of the scenarios discussed in \cref{sec:scenarios}.
Specifically, the event binding capability allows to utilize the managed blockchain instances as an event emitter (see \cref{sec:blockchains-as-event-emitters}).
Furthermore, the smart contract invocation capabilities allow to treat smart contracts as serverless functions (see \cref{subsec:scenario-3}).
However, since no integration with an orchestration service is provided, and because of the limited capabilities to communicate with external nodes, scenarios involving function orchestration or cross-platform integration are not supported (see \cref{fig:orchestration,sec:integration}).

\subsubsection{Amazon Web Services}
Amazon Managed Blockchain~\cite{aws:baas} is the BaaS offering of the AWS cloud platform.
It provides the capabilities to create and manage Hyperledger Fabric blockchain networks (see \req{A-1}).
A single instance can incorporate members of multiple AWS accounts, each with the ability to provision peer nodes.
These nodes run exclusively within the instantiated blockchain network, i.e., the addition of external nodes is not supported (see \req{A-4}).

AWS provides a set of tools and services\footnote{For details: \url{https://docs.aws.amazon.com/managed-blockchain/latest/managementguide/security-iam.html}} to manage authentication and authorization tasks that ensure secure access to the managed blockchain instance~(see \req{A-2}).
The offering does not directly allow the deployment and management of chaincodes (smart contracts of Hyperledger Fabric).
However, it facilitates external access to the instance resources, such as the peer nodes, by creating an interface Virtual Private Cloud (VPC) endpoint.
This allows authorized external applications and client machines to access the Hyperledger Fabric CLI and SDK, which enables, among other things, to request the installation, instantiation and upgrade of chaincodes (see \req{C-2}).
Furthermore, functions hosted on AWS Lambda, the FaaS offering of AWS, can trigger chaincode functions directly via the Hyperledger Fabric SDK (see \req{B-2}).
However, no native Lambda function trigger for blockchain events exists (see \req{B-1}), but this can be achieved indirectly via a combination of general-purpose services as discussed in~\cref{sec:blockchain-awareness}.
AWS Managed Blockchain claims future support for the public Ethereum blockchain (see \req{A-4}).
\req{B-3} would then need to be carefully evaluated for fulfillment, since blockchain forking is an issue with most public blockchains.
Finally, the limited ability to monitor blockchain events and activate serverless components based on them means that only one scenario is supported, and only partially (see \cref{subsec:scenario-3}).

\subsubsection{Cloudflare}
Cloudflare provides the Cloudflare Distributed Web service~\cite{Cloudflare2020DistributedWeb}, which includes a managed Ethereum~\cite{wood2018ethereum} gateway that allows authorized clients to interact with the public Ethereum blockchain (Ethereum \textit{Mainnet}) (see \req{A-4}).
The gateway exposes a subset of the standard Ethereum API via a JSON-RPC protocol~\cite{ethereum2019JsonRpc}.
The provided operations allow, among other things, to deploy pre-compiled smart contracts (see \req{C-2}), as well as to invoke smart contract functions (see \req{B-2}).
However, the live monitoring of smart contract events is not allowed (see \req{B-1}).
This hinders the ability to introduce Oracles to support interacting with external services (see \req{B-4}).
On the other hand, functions of the Cloudflare Workers FaaS platform~\cite{Cloudflare2020Workers} can deploy smart contracts and invoke smart contracts via the standard Ethereum SDK (web3).
Nonetheless, since invoking a state-changing smart contract function requires submitting a pre-signed transaction, the serverless function needs to manage Ethereum account credentials itself (see \req{A-2}).
Furthermore, the platform does not ensure that the account used to sign the transaction has enough funds nor that the transaction becomes durably committed (see \req{A-3} and \req{B-3}).
While Cloudflare is unique in terms of (i)~providing a gateway rather than a BaaS offering and (ii)~supporting access to a public blockchain, its inability to monitor smart contract events means only partial support for one of the discussed scenarios (see \cref{subsec:scenario-3}).

\subsubsection{Huawei Cloud}
Huawei Cloud provides the Blockchain Service (BCS)~\cite{HuaweiCloud2020BCS}, a BaaS offering with a PaaS service model.
Like most other BaaS offerings, BCS allows the creation and management of Hyperledger Fabric~\cite{androulaki2018fabric} permissioned blockchain networks via a specialized management console (see \req{A-1}).
To facilitate the development of chaincodes, the console provides a set of downloadable chaincode templates (see \req{C-1}).
Furthermore, using the same console, chaincodes can be edited, compiled, deployed, and instantiated on network peers (see \req{C-2}).
To invoke a chaincode function, a client application needs to use the Hyperledger Fabric Golang SDK, and needs to manually sign request messages using a private key (downloadable from the management console)  that corresponds to a blockchain-identity (see \req{A-2}).
This also applies to functions of the FunctionGraph service~\cite{HuaweiCloud2020FunctionGraph}, a FaaS platform hosted by Huawei Cloud (see \req{B-2}).
Moreover, monitoring chaincode events can also be performed using the aforementioned SDK, i.e., the platform does not support triggering a chaincode function based on a chaincode event (see \req{B-1}).
Apart from the SDK, BCS provides a REST API capable of invoking chaincode functions but requires the invoking application to manage its own blockchain identity (private key + certificate).
Finally, BCS does not facilitate incorporating external nodes in the blockchain network (see \req{A-4}), nor does it not have the ability to actively trigger other services based on chaincode events (see \req{B-4}).
The low integration capabilities of the BCS BaaS, specifically, in terms of chaincode events, results in the fact that Huawei Cloud only guarantees partial support for one of the discussed scenarios~(see \cref{subsec:scenario-3}).

\subsubsection{IBM Cloud}
IBM Cloud offers two categories of managed blockchains.
First, enterprises can join existing consortia already deployed on the cloud platform, such as IBM Food Trust~\cite{ibm:food-trust}, and IBM Rapid Supplier Connect~\cite{ibm:rapid-supplier-connect}.
Second, the cloud platform also allows creating new permissioned blockchain instances that other members can join later.
In this category, referred to as IBM Blockchain Platform~\cite{ibm:baas}, the user has the option to either deploy the BaaS platform on multiple Kubernetes distributions on private, public, or hybrid clouds, or to deploy it on a single Kubernetes cluster fully managed by the IBM Cloud (see \req{A-1}).
In all cases, the supported blockchain is Hyperledger Fabric~\cite{androulaki2018fabric}.
To perform managerial tasks, like instantiating new blockchain networks, adding peers, and managing certificates and user roles, the platform provides both a GUI-based console, and an SDK (see \req{A-2}).
Moreover, the managed blockchain can incorporate external peer nodes (see \req{A-4}).
To develop, debug, and deploy chaincodes, i.e., Hyperledger Fabric smart contracts, one can use IBM Blockchain Platform Developer Tools which are available as a Visual Studio Code extension~\cite{ibm:blockchain-develop-vscode}.
The tool provides tutorials and sample contracts to lower the learning curve for interested developers (see \req{C-1}).
On the other hand, IBM Blockchain Platform does not provide sufficient integration with other cloud services.
Specifically, events emitted from the managed blockchain cannot be directly sent to other cloud-based or external services, nor can it trigger Cloud Functions, the serverless offering of IBM Cloud (see \req{B-1} and \req{B-4}).
Nonetheless, this task can be still achieved indirectly with a polling-based approach~\cite{Thomas2019IBMFunctionTriggering}.
Additionally, Cloud Functions can trigger chaincode functions directly via one of the supported SDKs (see \req{B-2}). 
In this case blockchain credentials are accessed using the IBM Cloud Identity and Access Management service~\cite{ibm:iam}.
Finally, the limited ability to monitor blockchain events and subsequently activate serverless components means that only one discussed scenario is supported, and only partially (see \cref{subsec:scenario-3}).

\subsubsection{Microsoft Azure}
Azure Blockchain Service~\cite{ms:baas} is the BaaS offering from Microsoft.
Azure Blockchain Service allows creating and managing instances of the ConsenSys Quorum permissioned blockchain (see \req{A-1}).
Azure identities are mapped to blockchain identities with the help of Azure Active Directory and a reverse proxy handling requests before reaching managed blockchain nodes, and roles in the consortium (Administrators and Users) can be controlled from the Azure Portal (see \req{A-2}).
Furthermore, using the Blockchain Development Kit\cite{ms:baas-devkit}, developers can create and deploy smart contracts to the BaaS instance (see \req{C-2}).
On the other hand, there are multiple integration possibilities between Azure Blockchain Service and Azure Functions, the FaaS offering from Azure.
For instance, a managed blockchain consensus event, or a smart contract event can be captured by the Blockchain Data Manager service and directed to the Event Grid service, which can, e.g., trigger serverless functions (\req{B-1}), or even an external service via a webhook (\req{B-4}).
Furthermore, an Azure Function can directly trigger the execution of a smart contract function by using the corresponding Quorum blockchain SDK (see \req{B-2}).
Even more interestingly, a connector exists that connects Ethereum-based blockchains including, public and permissioned instances, with the Azure Logic Apps service\cite{ms:logicapps-ethereum}.
Azure Logic Apps allows to create workflows that can automate multi-step tasks involving a variety of services.
Using the aforementioned connector, a workflow can receive smart contract events, and can invoke smart contract functions.
This is especially helpful since Logic Apps can be further integrated with Azure Functions, and many more services and APIs.
It is worth mentioning here that since the connector supports connecting with the public Ethereum blockchain, the tasks of ensuring the sufficiency of funds and durability of transactions (see \req{A-3} and \req{B-3}) become relevant, but the developer needs to take care of them manually since no explicit platform support exists.
Another integration option is via the Blockchain Workbench Service\cite{ms:azure-bc-workbench}, which is a layer on top of blockchains that eases their management and focuses on the rapid development of smart contracts with the help of pre-defined templates (see \req{C-1}), and on an easy two-way integration experience with blockchain-external applications and services via the usage of Azure Service Bus, and Event Grid.
Specifically, the workbench exposes a REST API that external applications can use to invoke blockchain management and smart contract-related operations.
With the help of a Transaction Builder and Signer component integrated with the Azure Key Vault service, these requests can be transformed into signed transactions of the format expected by the target blockchain.
A Distributed Ledger Technology (DLT) Watcher component monitors the underlying blockchains and can publish blockchain events to the Event Grid or the Service Bus.
A major benefit of this service is that it can be integrated with existing Azure Blockchain Service instances, as well as any accessible Proof-of-Authority (PoA)-based Ethereum instance~\cite{Szilagyi2017Clique,Barinov2018PoA} (see \req{A-4}).

Overall, considering the set of capabilities provided by the Azure Blockchain Service, the platform can achieve most of the scenarios discussed in \cref{sec:scenarios}.
Specifically, the event binding capability allows to utilize the managed blockchain instances as an event emitter (see \cref{sec:blockchains-as-event-emitters}).
Furthermore, the smart contract invocation capabilities allow to treat smart contracts as serverless functions (see \cref{sec:blockchains-as-event-emitters,subsec:scenario-3}).
Moreover, the ability to integrate smart contract invocations and events with the Logic Apps service, which can act as a serverless orchestrator, means that the scenario described in \cref{sec:scenario-function-orchestration} (case a) is supported.
Lastly, the ability to connect to external blockchain nodes, in addition to the ability to push events to an external API via webhooks ensure that the \textit{Blockchain as a Message Bus} scenario (see~\cref{sec:integration}) can be theoretically supported.
However, since no blockchain-based orchestrator is integrated, relevant scenarios are not supported (see \cref{sec:scenario-function-orchestration} (case b), and the \textit{Blockchain as Process Manager} scenario in \cref{sec:integration}).

\subsubsection{Oracle Cloud}
Oracle Blockchain Platform (OBP)~\cite{oracle:baas} is the BaaS offering from Oracle Cloud.
It allows to create and manage instances of the Hyperledger Fabric~\cite{androulaki2018fabric} blockchain on one or more Oracle Cloud Infrastructure data centers.
The service also allows incorporating blockchain nodes deployed on-premise, and third-party clouds (see \req{A-4}).
Similar to other BaaS offerings, OBP has a specialized web console that allows performing managerial tasks like instantiating new blockchain deployments, adding peers and users to the network, and manage certificates (see \req{A-1}).
OBP leverages built-in Oracle Identity service integration, which can be used to authenticate users and applications wishing to interact with hosted blockchains.
On the other hand, to develop new chaincodes, a Visual Studio Code extension called the Blockchain App Builder~\cite{oracle:bc-app-builder} is available.
This extension allows using a comprehensive set of sample chaincodes covering several potential use-cases. 
Furthermore, it facilitates debugging the developed smart contracts and directly deploying them to an existing blockchain instance (see \req{C-1} and \req{C-2}).
In terms of integration with other services and external applications, the platform provides multiple options (see \req{B-1}, \req{B-2}, and \req{B-4}).
For example, a REST API that covers both administrative and application operations (like subscribing to events and invoking chaincodes).
Moreover, authenticated applications, like Oracle Functions, can use the Hyperledger Fabric SDK to perform similar tasks.
Besides, the platform provides native integration for blockchain events and treats them according to the CloudEvents specification~\cite{CloudEvents2018}, which allows (i) incorporating them into Streaming Service streams, (ii) publishing them as Notification Service messages, which can be consumed by other services or pushed to external HTTP endpoints, and (iii) use them to trigger Oracle Functions.
Finally, OBP provides a set of plug-and-play adapters that can be used by the Oracle Integration Service to allow various Oracle SaaS, PaaS and on-premise applications to perform tasks like invoking chaincode functions, and subscribing to events.

Considering the set of capabilities provided by the OBP offering, the platform can achieve only two of the scenarios discussed in \cref{sec:scenarios}.
Specifically, the event binding capability allows to utilize the managed blockchain instances as an event emitter (see \cref{sec:blockchains-as-event-emitters}).
Furthermore, the smart contract invocation capabilities allow to treat smart contracts as serverless functions (see \cref{sec:blockchains-as-event-emitters,subsec:scenario-3}).
However, since no integration with an orchestration service is provided, and because of the limited capabilities to communicate with external nodes, scenarios involving function orchestration or cross-platform integration are not supported (see \cref{fig:orchestration,sec:integration}).


\setlength\tabcolsep{4pt}
\begin{table*}[t!]
    \centering
    \begin{threeparttable}[b]
        \centering
        \captionsetup{justification=raggedright,singlelinecheck=false,format=hang}
        \caption{Review of considered BaaS offerings with respect to the requirements formulated in~\Cref{sec:requirements}. In the table, we use the following abbreviations: \textsf{HF} stands for "Hyperledger Fabric", \textsf{BC} stands for "Blockchain", \textsf{SC} stands for "Smart Contract", \textsf{Eth} stands for "Ethereum", and "n/a" stays for \enquote{not specified}, meaning that the requirement is not applicable for the supported blockchain(s). Additionally, we use the symbol \ding{51} to denote that the respective BaaS offering fulfills the requirement and we use \ding{53} otherwise. \\ Moreover, we denote the event binding direction using uni- and bi-directional arrow signs~($\to$, $\leftrightarrow$).}
        \label{tab:baas-review}
        \scriptsize
        \def\arraystretch{1.5}
        \definecolor{lightgray}{gray}{0.9}
        \begin{tabularx}{.95\textwidth}
            {@{} 
                c 
                m{32mm}
                m{16.8mm} 
                m{16mm} 
                m{16mm} 
                m{16mm} 
                m{16mm} 
                m{17.5mm} 
                m{17mm} 
            }
            \hline
            
            \multicolumn{2}{l}{}
            & \centering \textit{Alibaba Cloud} 
            & \centering \textit{AWS} 
            & \centering \textit{Cloudflare}
            & \centering \textit{Huawei Cloud} 
            & \centering \textit{IBM Cloud} 
            & \centering \textit{Microsoft Azure} 
            & \centering \arraybackslash \textit{Oracle Cloud} \\
            \hline
            
            & \centering Supported Blockchain Service Type
            & \centering BaaS (PaaS) 
            & \centering BaaS (PaaS) 
            & \centering Gateway as a Service 
            & \centering BaaS (PaaS) 
            & \centering BaaS (PaaS), BaaS (SaaS) 
            & \centering BaaS (PaaS), Deployable Gateway 
            & \centering \arraybackslash BaaS (PaaS) \\  
            \hline
            
            \parbox[t]{3mm}{\multirow{4}{*}[-1.5em]{\rotatebox[origin=c]{90}{\begin{minipage}{9.5em}\centering \textit{Req. Group A}\end{minipage}}}}
            
            & \cellcolor{lightgray} \centering Blockchain Node is a Serverless Component (\req{A-1})
            & \cellcolor{lightgray} \centering Quorum, HF, Ant BC 
            & \cellcolor{lightgray} \centering HF 
            & \cellcolor{lightgray} \centering Eth mainnet 
            & \cellcolor{lightgray} \centering HF 
            & \cellcolor{lightgray} \centering HF 
            & \cellcolor{lightgray} \centering Quorum, Eth PoA, Eth mainnet 
            & \cellcolor{lightgray} \centering \arraybackslash HF \\  
            
            & \centering Access to Authorized Blockchain Account (\req{A-2})
            & \centering RAM service manages credentials 
            & \centering Application manages BC credentials 
            & \centering Application manages BC credentials 
            & \centering Application manages BC credentials 
            & \centering IAM service manages credentials 
            & \centering AKV service manages credentials 
            & \centering \arraybackslash OIM service manages credentials \\  
            
            & \cellcolor{lightgray} \centering Sufficient Funds for Transaction Processing (\req{A-3})
            & \cellcolor{lightgray} \centering n/a 
            & \cellcolor{lightgray} \centering n/a 
            & \cellcolor{lightgray} \centering Ensured by application 
            & \cellcolor{lightgray} \centering n/a 
            & \cellcolor{lightgray} \centering n/a 
            & \cellcolor{lightgray} \centering Ensured by application 
            & \cellcolor{lightgray} \centering \arraybackslash n/a \\  
            
            & \centering Access to External Blockchain Nodes (\req{A-4})
            & \centering \ding{51}\tnote{(1)} 
            & \centering \ding{51}\tnote{(2)} 
            & \centering \ding{51} 
            & \centering \ding{53} 
            & \centering \ding{51} 
            & \centering \ding{51} 
            & \centering \arraybackslash \ding{51} \\  
            \hline            
            
            \parbox[t]{3mm}{\multirow{4}{*}[-4.5em]{\rotatebox[origin=c]{90}{\begin{minipage}{9.5em}\centering \textit{Req. Group B}\end{minipage}}}}
            
            & \cellcolor{lightgray} \centering Support for Blockchain Events Subscription (\req{B-1})
            & \cellcolor{lightgray} \centering \ding{51} 
            & \cellcolor{lightgray} \centering \ding{53} 
            & \cellcolor{lightgray} \centering \ding{53} 
            & \cellcolor{lightgray} \centering \ding{53} 
            & \cellcolor{lightgray} \centering \ding{53} 
            & \cellcolor{lightgray} \centering \ding{51} 
            & \cellcolor{lightgray} \centering \arraybackslash \ding{51} \\  
            
            & \centering Support for Smart Contract Invocation (\req{B-2})
            & \centering Function Compute (via REST API) 
            & \centering AWS Lambda (via BC SDK) 
            & \centering Workers (via BC SDK) 
            & \centering Function Graph (via BC SDK + REST API) 
            & \centering IBM Cloud Functions (via BC SDK) 
            & \centering Azure Functions (via BC SDK + REST API), Azure Logic Apps (via a connector) 
            & \centering \arraybackslash Oracle Functions (via BC SDK + REST API), other services (via adapters) \\  
            
            & \cellcolor{lightgray} \centering Transaction Durability Guarantees (\req{B-3})
            & \cellcolor{lightgray} \centering n/a 
            & \cellcolor{lightgray} \centering n/a 
            & \cellcolor{lightgray} \centering \ding{53} 
            & \cellcolor{lightgray} \centering n/a 
            & \cellcolor{lightgray} \centering n/a 
            & \cellcolor{lightgray} \centering \ding{53} 
            & \cellcolor{lightgray} \centering \arraybackslash n/a \\  
            
            & \centering Blockchain as Active Communicator (\req{B-4})
            & \centering RocketMQ pushes events to external API 
            & \centering \ding{53} 
            & \centering \ding{53} 
            & \centering \ding{53} 
            & \centering \ding{53} 
            & \centering EventGrid pushes events to external HTTP API via webhook 
            & \centering \arraybackslash Notification Service pushes events to external HTTP API \\  
            \hline
            
            \parbox[t]{3mm}{\multirow{2}{*}[-.5em]{\rotatebox[origin=c]{90}{\begin{minipage}{9.5em}\centering \textit{Req. Group C}\end{minipage}}}}
            
            & \cellcolor{lightgray} \centering Support for Smart Contracts Development (\req{C-1})
            & \cellcolor{lightgray} \centering \ding{51} 
            & \cellcolor{lightgray} \centering \ding{53} 
            & \cellcolor{lightgray} \centering \ding{53} 
            & \cellcolor{lightgray} \centering \ding{51} 
            & \cellcolor{lightgray} \centering \ding{51} 
            & \cellcolor{lightgray} \centering \ding{51} 
            & \cellcolor{lightgray} \centering \arraybackslash \ding{51} \\  
            
            & \centering Support for Deployment Automation (\req{C-2})
            & \centering SC deployment, SC event $\to$ serverless component binding 
            & \centering SC deployment 
            & \centering SC deployment 
            & \centering SC deployment 
            & \centering SC deployment 
            & \centering SC deployment, SC event $\to$ serverless component binding, orchestrator $\leftrightarrow$ SC binding 
            & \centering \arraybackslash SC deployment, SC event $\to$ serverless component binding, serverless event $\to$ SC binding \\  
            \hline
            
        \end{tabularx}
        \begin{tablenotes}
            \scriptsize
            \item[(1)] \textit{Only true for ConsenSys Quorum}
            \item[(2)] \textit{Will be fulfilled when public Ethereum is supported}
        \end{tablenotes}
    \end{threeparttable}
\end{table*}

\subsection{Technology Comparison Highlights}
\Cref{tab:baas-review} summarizes the results of the conducted technology survey with respect to the readiness of the considered platforms to fulfill the implementation requirements formulated in~\Cref{sec:requirements}.
Firstly, the majority of providers offer the blockchain access in the form of a BaaS offerings~(6/7), whereas the gateway option is noticeably rare~(2/7).
Moreover, the BaaS offerings typically follow PaaS cloud service model's style, meaning that users can design the blockchain network, configure it (including creation of certificates), and instantiate it using a UI instead of Docker compose files or other means of automation.
It is worth highlighting that in such mode, blockchain node access happens no in purely-serverless fashion, since the entire network has to be run with even when it is not used and additional management efforts are needed.

One can notice several interesting patterns in the providers' preferences towards the supported blockchain systems.
Most reviewed offerings~(5/7) only support one blockchain system, with the majority choosing to support Hyperledger Fabric~\cite{androulaki2018fabric}~(4/7).
Moreover, the overall focus on permissioned blockchains is obvious: only one provider~(Cloudflare) does not support any permissioned blockchains, whereas permissionless blockchains~(only Ethereum) are only supported by Azure and Cloudflare.

The next important observation is that the majority of reviewed providers offer integration with other services in a basic form, as shown by fulfillment of requirements from group B in~\Cref{tab:baas-review}).
For example, reviewed BaaS offerings provide limited options for binding events \textit{from} smart contracts \textit{to} serverless components, and limited integration with function orchestrators.
Similar with function invocation, which is only supported using blockchain SDKs.
Moreover, the option to use blockchain as an active communicator~(see \req{B-4}) is supported only by Microsoft Azure and Oracle, which is implemented by pushing events from an event aggregation service to external APIs.

Finally, in terms of improving the smart contract development process, the majority of providers~(5/7) offer some sort of support.
Moreover, the deployment automation options are rather rudimentary, i.e., most offerings provide means for automating smart contracts deployment, typically by providing SDKs and exposing APIs that enable, e.g., defining imperative-style deployment automation scripts.
Declarative deployment options that can cover serverless offerings with blockchain-specific components are not present.
Moreover, some options for automating the deployment of event bindings is offered by three providers, who focus on binding smart contract events to serverless components~(3/3), and binding cloud events and smart contracts~(2/3).

To summarize, we point out that \textit{Scenario 2: Smart Contracts as Serverless Functions} (see \cref{subsec:scenario-3}) is achieved by all (7/7) providers, whereas \textit{Scenario 1: Blockchains as Serverless Event Emitters} (see \cref{sec:blockchains-as-event-emitters}) is only achieved by Alibaba Cloud, Microsoft Azure, and Oracle Cloud (3/7) since this scenario requires having the ability to bind a blockcahin event to serverless components, which is only achieved by these providers.
Moreover, \textit{Scenario 3: Blockchains for Function Orchestration - case a} (see \cref{sec:scenario-function-orchestration}), and \textit{Scenario 4: Blockchains as Facilitators for the Integration of Serverless Applications - Blockcahins as Message Bus} (see \cref{sec:integration}) can only be supported by Microsoft Azure (1/7) since it is the only provider that allows to bind smart contract functions to a function orchestrator service.
Finally, \textit{Scenario 3: Blockchains for Function Orchestration (case b)} (see \cref{sec:scenario-function-orchestration}), and \textit{Scenario 4: Blockchains as Facilitators for the Integration of Serverless Applications - Blockcahins as Process Manager} (see \cref{sec:integration}), cannot be achieved by any of the considered providers (0/7) since none of them supports the integration of a blockchain-based process engine, such as Caterpillar~\cite{Lopez-Pintado2017}.

\section{Major Findings}
\label{sec:blockchain-awareness}

In this subsection, we highlight the main findings of this work and discuss several interesting issues that can be used a basis for future research.

\subsection{Blockchain-awareness of Infrastructures}

When analyzing how to employ existing blockchains in serverless contexts, one needs to note that the implementation of certain requirements formulated in~\Cref{sec:requirements} highly depends on the chosen set of serverless component offerings and the level of their support for integration with blockchains, i.e., how aware the chosen provider's infrastructure of blockchains is.
On the one hand, the chosen provider's infrastructure might be completely \textit{blockchain-ignorant}, e.g., its serverless offerings do not have any native blockchain integration options and thus no standard way to use blockchain events and smart contracts as part of the application.
On the other hand, the chosen infrastructure might be \textit{blockchain-aware} meaning that it is possible to use available blockchain integration features offered as part of blockchain-centric or general-purpose services for implementing the discussed scenarios (see~\cref{sec:scenarios}).
Obviously, here we mean a spectrum of blockchain-awareness rather than a strict yes/no answer, as providers offer varying sets of features for different blockchain systems as shown in~\Cref{sec:system-support}.
Although many of the prominent public cloud providers offer some sort of blockchain-specific services, as shown in~\Cref{sec:system-support}, the sets of features vary significantly, influencing the \enquote{degree of blockchain-awareness}.


\subsection{Blockchain Access Options and Influence on Trust}
Much of the discussion around the usefulness of the blockchain technology, even the discussion we employed in this paper, revolves around the benefits of disintermediation, i.e., how limiting the role of trusted third-parties or their complete removal enhances the trust guarantees of the systems and promotes previously unforeseen use-cases.
Nonetheless, Baas is the most common option for blockchain node access services (see \req{A-1}) offered by the public cloud providers discussed in this survey, although it involves the management of all or most of blockchain network by the provider itself.
The influence BaaS has on the trust guarantees of the underlying blockchain system is discussed by Singh an Michels~\cite{Singh2018BaaSTrust}.
They argue that BaaS offerings are not identical, and that a balance between the ease of management and trust exists:
on one end of the spectrum are offerings that are totally managed, i.e., BaaS using the SaaS service model in which the client is not aware of any specifics regarding the blockchain system including its own blockchain identity.
Such services allow the provider to have full control over the network, the history of submitted transactions, whether to allow specific requests to go through or not, and even issuing requests using the client's credentials.
On the other end of the spectrum are BaaS offerings that use the IaaS service model, in which the tenants are offered with basic compute, storage, and network services and they need to setup instantiate, and manage the whole blockchain network themselves (certain third-party Azure ARM templates fall into this category).
In between these two extremes are BaaS offerings that operate under the PaaS service model.
The degree of trust guaranteed by these offerings depends on the division of management responsibilities between the provider and the tenant.
For example, if the credentials of blockchain accounts are managed by the provider (see \req{A-2}) and are only mapped to the cloud identity of the corresponding client applications, the provider is able to have full control over these accounts, including submitting transactions on their behalf.
However, if client applications have to manage their own credentials, the offering becomes \enquote{less managed}, but this enhances its trust properties.
Other similar management-trust trade-offs include the ability to construct blockchain networks that incorporate on-premise nodes, or nodes on different providers (hyper and multi clouds resp.).

Apart from traditional BaaS offerings, other options exist for serverless applications (and cloud applications in general) to access the blockchain network (all having different choices regarding the management-trust trade-off):
One option is to use a provider-hosted blockchain gateway, as is the case with Cloudflare~\cite{Cloudflare2020DistributedWeb}, which is a blockchain node part of a larger blockchain network (usually permissionless).
In this case, client applications manage their own credentials, and can verify independently, e.g., using other gateways, that their intended interactions with the blockchain are actually carried out honestly by the gateway, thus limiting the degree in which it can maliciously influence these interactions.
Another option is to use a third-party blockchain gateway like Infura~\cite{infura:infura}.
This separates the party that manages the blockchain node access from the party managing the cloud application, thus decreasing the probability of malicious acts targeting specific applications.
A more hypothetical example is if a blockchain node can be used as an internal part of  general-purpose serverless component type, e.g., a specialized blockchain-aware FaaS platform that transparently enables interaction with blockchain events and smart contracts.
Finally, one possibility is to set up a blockchain node on a provider-managed Kubernetes cluster relaxing the definition of serverless-ness.
Such pre-configured Kubernetes clusters can then be used for combining Kubernetes-based FaaS platforms such as OpenFaaS with the chosen blockchain system.

Overall, we see that the question of the degree of trust expected from a blockchain node access approach has no strict (yes/no) answer, but falls on a spectrum of options affected by the chosen management-trust trade-offs.

\subsection{Options for Blockchain Event Binding}
Since the requirements from group B discussed in~\Cref{req:group-B} are related to the roles a blockchain system could play in serverless applications, the degree of blockchain-awareness also highly influences the implementations options.
One good example is the implementation of event bindings, i.e., how to bind serverless components, such as FaaS-hosted functions, with blockchain events. 
The way such a binding can be implemented depends on the capabilities of the underlying infrastructure, e.g., a native integration between the BaaS and FaaS offering is offered. 
Another option is to use a combination of several serverless components that polls for blockchain events and triggers FaaS-hosted functions based on these events.
As shown in~\Cref{sec:baas}, an example of real-world native integration is a combination of Azure's BaaS offering with Azure Event Hub and Azure Functions, which, for example, enables implementing \textit{blockchain as event emitter} scenario discussed in~\Cref{sec:blockchains-as-event-emitters}.
This, of course, applies only to blockchains supported by Azure's BaaS offering.

\subsection{Options for Smart Contract DevOps and Invocation}
Similar considerations apply to the implementation of smart contracts invocation, i.e., an ideal case is a blockchain-aware FaaS platform which can handle smart contracts invocation based on the occurrence of events emitted by provider's general-purpose offerings.
As before, such functionality can, in theory, be also implemented by composing several serverless components together.
For example, as discussed in~\Cref{sec:baas}, Azure supports such integration by combining Blockchain Workbench's REST API and Azure Key Vault to enable issuing signed transactions for a given blockchain.
This allows using FaaS-hosted functions triggering smart contracts, which however introduces additional component in the application topology that are solely responsible for enabling such integration.
The difference in implementation options also affects which logic has to be stored in smart contracts themselves, e.g., depending on the target service offering the output format of a smart contract's function has to conform to a specific schema.
An automatic deployment described in the \req{C-2}~\Cref{req:group-C} also highly depends on the deployment automation features.
As a result, requirements from group C~(\Cref{req:group-C}) are also affected by how blockchain-aware the chosen infrastructure of service offerings is.


\section{Related Work}
\label{sec:related-work}

To the best of our knowledge there are no publications neither trying to analyze how blockchains and smart contracts can be used in serverless architectures, nor formulating the underlying implementation requirements and investigating which existing technologies can enable such scenarios.
Nevertheless, the idea to use blockchains and smart contracts in serverless architectures appears in several research works, which we discuss in the following.

In principle, the topic of serverless computing is relatively new and no fixed boundaries for available component types are defined, which can also be seen in several publications highlighting that new component types might be introduced in the future~\cite{related:motivation:baldini2017serverless, serverless:jonas2019cloud}.
Some researchers discuss the concepts of blockchains and smart contracts from the perspective of reduced infrastructure management or even explicitly call them serverless.
The decentralzied and distributed nature of blockchains and smart contracts is one of the main reasons to link them with serverless infrastructures~\cite{related:motivation:Kuehn2019,related:motivation:KADAM2020183,related:motivation:riesco2020cybersecurity}.
In certain cases, the processing of smart contract's function invocations is described as being similar to the FaaS processing model~\cite{related:motivation:kfir2019daml}.
Moreover, some blockchain-focused papers try designing systems that combine blockchains and FaaS-based architectures to realize blockchain-related operations using more serverless-style components~\cite{related:motivation:weber2018blockchain,related:BC-first:chen2018fbaas,related:BC-first:bumblauskas2020blockchain,related:BC-first-hybrid-serverful:benedict2020serverless,related:BC-first:kaplunovich2019scalability,related:BC-first:stanke2020development}.
For example, Kaplunovich et al.~\cite{related:BC-first:kaplunovich2019scalability} use a serverless AWS-based architecture to benchmark blockchains, benefiting from the simplified interaction with the provider-managed offerings such as AWS Lambda and AWS DynamoDB.
Stanke et al.~\cite{related:BC-first:stanke2020development} introduce a hybrid architecture that combines the IOTA blockchain system with a serverless part that relies on an AWS service that offers finite element simulation.
Another blockchain-centric example proposed by Chen and Zhang~\cite{related:BC-first:chen2018fbaas} uses a FaaS platform as an internal component for a lighter implementation of a BaaS offering.

Even more interesting scenarios of combining blockchains and serverless architectures for our work are those focusing more on the serverless part of the equation.
Oh and Kim~\cite{related:SC-FaaS:oh2019serverless} outline the idea of using serverless, FaaS-based architectures as means to offload the computation from blockchains.
In this scenario, the AWS Lambda-based part of the system serves as a cheaper and more efficient substitute for blockchain's compute resources, allowing to ship computation \textit{from blockchain smart contracts to FaaS}.
One more serverless-centric scenario by Ghaemi et al.~\cite{related:SC-FaaS:ghaemi2020chainfaas} focuses on offloading computation from cloud-hosted FaaS platforms to unused compute resources of personal computers belonging to any user willing to participate in such a scheme.
Focusing on green computing, authors introduce a specialized, blockchain-based FaaS platform that supports running user-provided code on personal computers of network's participants who, as an incentive, receive a percentage of the transaction for serving requests.
Another crossover between blockchains and serverless is described by Tonelli et al.~\cite{related:SC-FaaS:tonelli2019implementing}.
The authors implement a microservice-based architecture~(MSA) using solely smart contract functions as substitutes for traditional processing components, e.g., using PaaS.
The discussed architecture reflects the serverless API use case~\cite{serverless:baldini2017serverless,Yussupov2019_Portability} and can be implemented in a serverless manner, making it an example of shipping computation \textit{from FaaS to blockchain smart contracts}.

While each discussed publication tries to combine blockchains with the notion of serverless~(mainly focusing on FaaS) in different ways, the overall goal is typically to solve a particular task, e.g., benchmarking, or implementing a specialized BaaS or FaaS offering, none of the aforementioned works neither aims at classifying the roles blockchains and smart contracts can play in serverless architectures, nor focuses on the aspects of serverless architecture design per se.
Moreover, none of these works analyzes the underlying implementation requirements and/or investigates which industrial solutions can enable such blockchain usage scenarios.
\looseness=-1


\section{Conclusion and Future Work}
\label{sec:conclusion}

In this paper, we analysed which distinct roles blockchains and smart contracts can play in serverless architectures and to what extent current public clouds enable such integration of blockchains.
More specifically, we identified a set of scenarios focusing on distinct tasks that can be accomplished by blockchains and smart contracts.
This includes the use of (i)~blockchain events in serverless architectures, e.g., to trigger FaaS functions, (ii)~smart contracts as a substitute for FaaS functions, (iii)~blockchains in the context of function orchestration, and (iv)~blockchains as integration facilitators for multiple serverless architectures in single- and  multi-cloud scenarios.
The results of reviewing the existing blockchain services from public cloud providers that also have FaaS platforms offerings showed that at least some level of support for using blockchains in serverless architectures is introduced by each provider, although the degree of native blockchain integration support varies significantly. 

Offerings that are more \textit{blockchain-aware}, such as Azure Blockchain Service, Oracle Blockchain Platform, and Alibaba Cloud BaaS enable the implementation of many of the discussed scenarios for the supported blockchain systems.
Moreover, among the considered providers, Azure's ecosystem of services has the best integration means for using blockchains and smart contracts as part of serverless applications.
Finally, some of the identified scenarios are currently not supported by providers, since no blockchain-based process engine services are offered by them.

On the other hand, further interesting findings can be inferred from this study: for example, the degree of \textit{blockchain-awareness} of cloud infrastructures is a continuum of values rather than a binary value, and it directly influences the ability to realize the identified scenarios.
Furthermore, the way a platform allows hosted applications to access the blockchain network greatly affects the degree of trust guaranteed.
Basically, a trade-off exists between the ease of management of blockchain access and the degree of trust guaranteed by the solution.

The main focus of this work was to highlight multiple architectural decisions that come into play when using blockchains in serverless architectures and emphasize the strong sides of such integration.
In future work, we plan to focus on more practical aspects of such integration scenarios, such as how to enable deployment automation using existing deployment systems, as well as how to support developers with the decision making process when designing serverless architectures incorporating blockchains and smart contracts as application components. 

\medskip \noindent
\textbf{Acknowledgements.}
This work is partially funded by the European Union's Horizon 2020 research and innovation project \emph{RADON}~(825040).

\bibliography{ms}

\smallskip \noindent
{\footnotesize All links were last followed on \today.}

\end{document}